\documentclass[preprint,journal]{IEEEtran}
\usepackage{graphicx}
\usepackage{subfigure}
\usepackage{amsmath}
\usepackage{color}
\usepackage{cite}
\usepackage{amsmath,amssymb,amsfonts}
\usepackage{algorithm}
\usepackage{algorithmic}
\usepackage{textcomp}
\usepackage{booktabs}
\usepackage{epstopdf}
\usepackage{array}
\usepackage{pgffor}
\usepackage{threeparttable}
\usepackage{multicol}
\usepackage{mathtools}
\usepackage[marginal]{footmisc}

\usepackage{hyperref}
\usepackage{cleveref}

\newcommand{\fig}[1]{\mbox{Figure~\ref{#1}}}
\newcommand{\eq}[1]{\mbox{Eq.~(\ref{#1})}}

\newcommand{\tab}[1]{\mbox{Table~\ref{#1}}}

\graphicspath{{./figure/}}
\begin{document}

\title{Neural Born Iterative Method For Solving Inverse Scattering Problems: 2D Cases}

\author{\IEEEauthorblockN{Tao Shan,~\IEEEmembership{Member,~IEEE}, Zhichao Lin, Xiaoqian Song, Maokun Li,~\IEEEmembership{Senior Member,~IEEE}, \\
		Fan Yang,~\IEEEmembership{Fellow,~IEEE}, and Shenheng Xu,~\IEEEmembership{Member,~IEEE}}
		\thanks{\color{red}This preprint has been published in IEEE Transactions on Antennas and Propagation on 01 November 2022. Please cite the final published version as [T. Shan, Z. Lin, X. Song, M. Li, F. Yang and S. Xu, "Neural Born Iterative Method for Solving Inverse Scattering Problems: 2D Cases," in IEEE Transactions on Antennas and Propagation, vol. 71, no. 1, pp. 818-829, Jan. 2023, doi: 10.1109/TAP.2022.3217333.]. The link is \url{https://ieeexplore.ieee.org/document/9934007}.}
		\thanks{\color{red}Digital Object Identifier 10.1109/TAP.2022.3217333}
	\thanks{This work was supported in part by the Institute for Precision Medicine, Tsinghua University, in part by the National Natural Science Foundation of China under Grant 61971263 and 62171259, the Biren Technology, the BGP Inc, and the China Postdoctoral Science Foundation under Grant 2022M711764.}
	\thanks{Tao Shan, Zhichao Lin, Maokun Li, Fan Yang and Shenheng Xu are with Beijing National Research Center for Information Science and Technology (BNRist), Department of Electronic Engineering, Tsinghua University, Beijing 100084, China. (e-mail:maokunli@tsinghua.edu.cn).
		
		Xiaoqian Song is with National Institute of Metrology, Beijing, 100013, China}}
\IEEEoverridecommandlockouts
\IEEEpubid{\makebox[\columnwidth]{0018-926X \copyright 2022 IEEE \hfill} \hspace{\columnsep}\makebox[\columnwidth]{ }}
\maketitle
\IEEEpubidadjcol
\begin{abstract}
	In this paper, we propose the neural Born iterative method (NeuralBIM) for solving 2D inverse scattering problems (ISPs) by drawing on the scheme of physics-informed supervised residual learning (PhiSRL) to emulate the computing process of the traditional Born iterative method (TBIM).
	NeuralBIM employs independent convolutional neural networks (CNNs) to learn the alternate update rules of two different candidate solutions regarding the residuals.
	Two different schemes are presented in this paper, including the supervised and unsupervised learning schemes.
	With the data set generated by the method of moments (MoM), supervised NeuralBIM are trained with the knowledge of total fields and contrasts.
	Unsupervised NeuralBIM is guided by the physics-embedded objective function founding on the governing equations of ISPs, which results in no requirement of total fields and contrasts for training.  
	Numerical and experimental results further validate the efficacy of NeuralBIM.
\end{abstract}
\vskip0.5\baselineskip
\begin{IEEEkeywords}
	Inverse scattering problem, Born iterative method, deep learning, supervised learning, unsupervised learning
\end{IEEEkeywords}
\IEEEpeerreviewmaketitle
\section{Introduction}
Inverse scattering are an ubiquitous tool to determine the nature of unknown scatterers with knowledge of scattered electromagnetic (EM) fields\cite{chen2018computational}, which has been applied widely across nondestructive testing\cite{salucci2016real}, biomedical imaging\cite{abubakar2002imaging}, microwave imaging\cite{pastorino2018microwave}, geophysical exploration\cite{abubakar20082, salucci2016multifrequency}, etc.
The intrinsic nonlinearity and ill-posedness pose long-standing challenges to ISPs\cite{chen2018computational}.
Many efforts are devoted to performing reliable inversion by addressing these two challenges, yielding noniterative and iterative inversion methods.
Noniterative inversion methods linearize ISPs under specific conditions, such as Born or Rytov approximation method\cite{slaney1984limitations, devaney1981inverse}, back-propagation (BP) method\cite{belkebir2005superresolution}, etc.
Iterative inversion methods transform ISPs into optimization problems of which optimal solutions are identified via an iterative process, for example, Born or distorted Born iterative method\cite{wang1989iterative, chew1990reconstruction}, contrast source inversion method\cite{van2001contrast, zakaria2010finite}, subspace-based optimization\cite{chen2009subspace}, level set method\cite{dorn2006level}, etc.
Prior information can be employed as a regularization to constrain and stabilize the iterative method for better inversion performance\cite{mojabi2009overview}.
Commonly applied regularizations include Tikhonov\cite{golub1999tikhonov}, total variation (TV)\cite{van1995total} and multiplicative regularizations\cite{van2001contrast}.
Besides, the compressive sensing paradigm can introduce an important type of regularization by decomposing the ISP solution based on a set of properly chosen basis functions\cite{candes2008introduction, oliveri2017compressive, guo2015microwave, oliveri2019compressive, pan2012compressive}.
Noniterative inversion methods hold good computational efficiency but limited applicability, while iterative ones can perform reliable inversion but suffer from immense computational loads.

\begin{figure}
	\centering
	\includegraphics[width=0.49\linewidth]{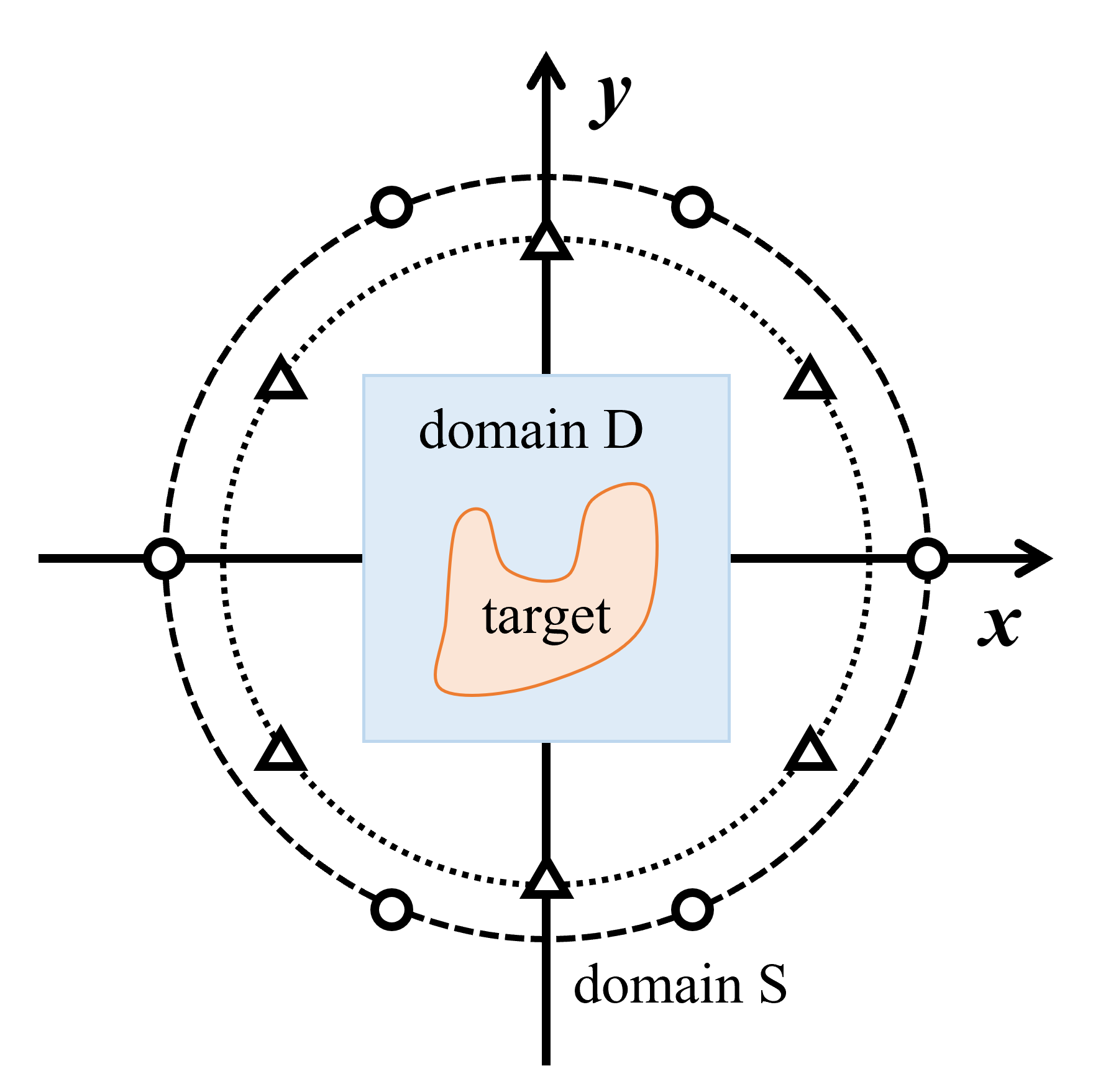}
	\caption{Model setup of inverse scattering problems. The target is inside the domain $D$ and the triangles and circles represent transmitters and receivers respectively.}
	\label{modelsetup}
\end{figure}
Recently, machine learning (ML), especially deep learning (DL), has been applied in the EM field, such as forward modeling\cite{guo2020application, shan2021application, shan2020study}, array antenna design\cite{massa2019dnns, shan2020coding, shan2021phase} and ISPs\cite{chen2020review,salucci2022artificial}, etc., which demonstrates unprecedented numerical precisions and computational efficiency\cite{chen2020review}.
The supervised descent method (SDM) is trained offline to learn and store the descent directions to guide the online inversions\cite{guo2019supervised}.
The powerful learning capacity of deep neural networks (DNNs), like CNNs\cite{xu2020deep,zhang2022accelerating}, generative adversarial networks\cite{song2021electromagnetic}, etc., is leveraged to retrieve the properties of scatterers from the scattered fields directly.
Moreover, the combination of DL and traditional ISP methods can achieve improved computational efficiency and inversion quality\cite{chen2020review}.
DNNs are trained to enhance the initial guess of inversions generated by noniterative inversion methods\cite{li2018deepnis, guo2021complex,wei2018deep}.
Conversely, iterative inversion methods can take as input the reconstructions predicted by DNNs\cite{chen2019learning,sanghvi2019embedding}.
Despite the remarkable success, most of the works mentioned above regard DNNs as "black-box" approximators, and the challenge of designing DNNs effective for ISPs has not been fully addressed.

The ordinary or partial differential equations (ODEs/PDEs) theory enables better insights into DNN properties, which guides the design of effective DNN architectures with better robustness and interpretability.
In \cite{lu2018beyond}, ResNet, PolyNet, FractalNet, and RevNet are linked to different numerical methods of ODEs or PDEs, respectively.
ResNet can also be interpreted from the perspective of dynamical systems\cite{haber2017stable, weinan2017proposal} or combined with the fixed-point iterative method\cite{shan2021physics}.
The spectral method and random matrix theory are combined to weigh the trainability of DNN architectures by analyzing their dynamical isometry\cite{xiao_dynamical_2018, tarnowski_dynamical_2019}. 
PDE-Net is proposed based on the similarities between finite difference operators and convolution operations\cite{long2018pde}.
New structures of ResNets for image processing are motivated by parabolic, and hyperbolic PDEs \cite{ruthotto2020deep}.

In this paper, we propose the neural Born iterative method by drawing on PhiSRL to emulate the computational process of the traditional Born iterative method.
Combining ResNet\cite{he2016deep} and the fixed-point iterative method, PhiSRL iteratively modifies a candidate solution until convergence by applying CNNs to predict the modification.
NeuralBIM builds the learned parameterized functions for the alternate update process of TBIM by drawing on the idea of PhiSRL.
Therefore, NeuralBIM bears the same computation as TBIM to reconstruct the total fields and scatterers simultaneously.
The supervised and unsupervised learning schemes of NeuralBIM are demonstrated and validated in this paper.
Supervised NeuralBIM is first presented to be trained by applying MoM to generate total fields and contrasts as labels.
Unsupervised NeuralBIM is proposed by deriving the physics-embedded objective function from the governing equations of ISPs, which further gets rid of total fields and contrasts for training.
Numerical and experimental results verify the efficacy of NeuralBIM.

This paper is organized as follows. 
Section \uppercase\expandafter{\romannumeral2} formulates inverse scattering problems. 
Section \uppercase\expandafter{\romannumeral3} introduces traditional Born iterative method. 
Both supervised and unsupervised neural Born iterative methods are introduced in Section \uppercase\expandafter{\romannumeral4}. 
Section \uppercase\expandafter{\romannumeral5} demonstrates the numerical and experimental results of neural Born iterative method.
Finally, Section \uppercase\expandafter{\romannumeral6} draws  conclusions and outlooks.
\begin{figure*}
	\centering
	\includegraphics[width=0.95\linewidth]{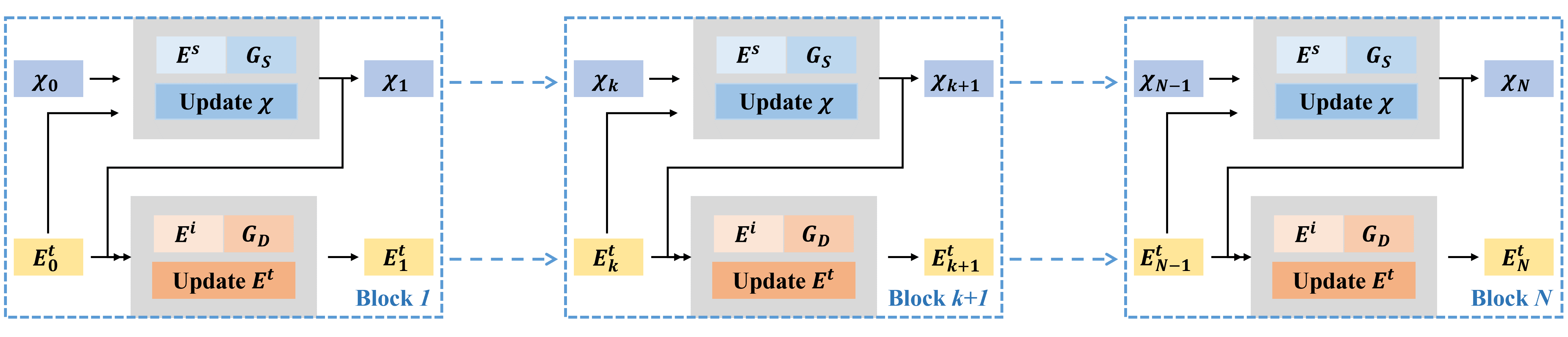}
	\caption{The architecture of Neural Born iterative method. NeuralBIM is assumed to have 7 blocks in this paper with Block $k$ denoting the $k$-th iteration. The contrast $\boldsymbol{\chi}$ is first updated and then employed to update the total field $\mathbf{E}^{t}$ in the $k$-th iteration. }
	\label{BIMNN}
\end{figure*}
\begin{figure}
	\centering
	\includegraphics[width=1\linewidth]{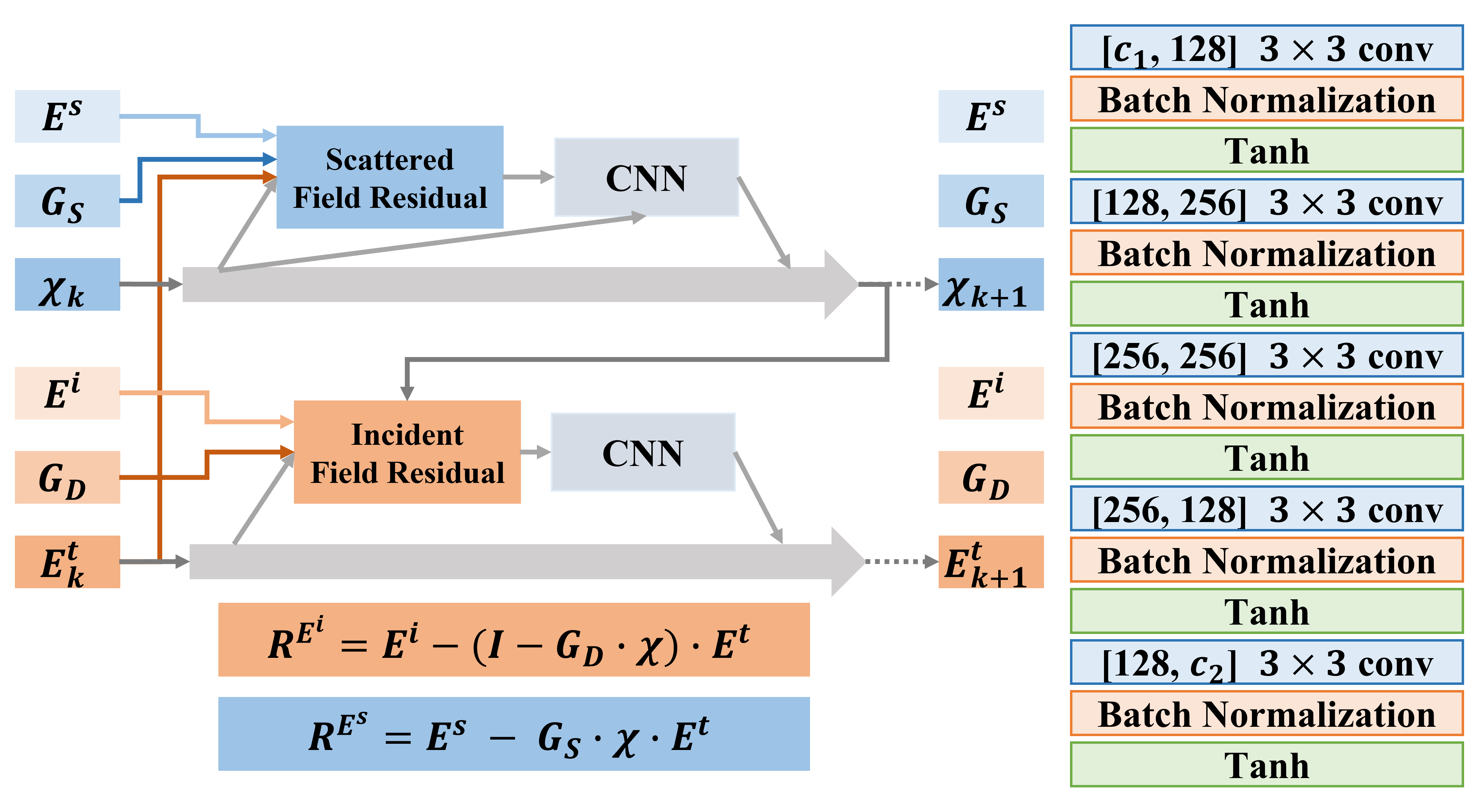}
	\caption{The iterative block of Neural Born iterative method. $\mathcal{R}_{k}^{\mathbf{E}^{i}}$ and $\mathcal{R}^{\mathbf{E}^{s}}_k$ represent the incident and scattered field residuals. $\boldsymbol{\chi}_{k+1}$ is updated by $\boldsymbol{\chi}$-CNN based on $\mathcal{R}^{\mathbf{E}^{s}}_k$ and $\boldsymbol{\chi}_{k}$. Then $\mathcal{R}_{k}^{\mathbf{E}^{i}}$ is obtained regarding $\boldsymbol{\chi}_{k+1}$ and then input to $\mathbf{E}^{t}$-CNN to update $\mathbf{E}_{k+1}^{t}$. The $\boldsymbol{\chi}$-CNN and $\mathbf{E}^{t}$-CNN share the same architecture but different parameter sets.}
	\label{BIMNNblock}
\end{figure}
\section{Inverse Scattering Problems}
In a 2D domain of interest (DoI) $D$, the transverse magnetic (TM) polarized line sources illuminate the unknown scatterers.
The incident field  $E^{i}$, total field  $E^{t}$ and scattered field  $E^{s}$ are related by
\begin{equation}
	E^{t}(\mathbf{r})=E^{i}(\mathbf{r})+k_b^2\int_{D} G_D \left(\mathbf{r}, \mathbf{r}^{\prime}\right) \boldsymbol{\chi}\left(\mathbf{r}^{\prime}\right) E^{t}\left(\mathbf{r}^{\prime}\right) d\mathbf{r}^{\prime}, \, \mathbf{r}  \in D 
	\label{eq1}
\end{equation}
\begin{equation}
	E^{s}(\mathbf{r}^{\prime \prime})=\int_{D} G_S \left(\mathbf{r}^{\prime \prime}, \mathbf{r}^{\prime}\right) \boldsymbol{\chi}\left(\mathbf{r}^{\prime}\right) E^{t}\left(\mathbf{r}^{\prime}\right) d\mathbf{r}^{\prime}, \, \mathbf{r^{\prime \prime}} \in S 
	\label{eq2}
\end{equation}
where $k_b$ is the wavenumber, $G_D$ and $G_S$ denote Green's functions in free space, $\boldsymbol{\chi}\left(\mathbf{r}\right)$ is the scatterer contrast and $S$ is the observation domain.
\eq{eq1} and \eq{eq2} can be discretized into a linear system of matrix equations by employing MoM:
\begin{equation}
	(\mathbf{I} - \mathbf{G}_D \boldsymbol{\chi}) \mathbf{E}^{t}  = \mathbf{E}^{i}
	\label{eq3}
\end{equation}
\begin{equation}
	\mathbf{E}^{s} = \mathbf{G}_S \boldsymbol{\chi} \mathbf{E}^{t}
	\label{eq4}
\end{equation}
ISPs aims to reconstruct $\boldsymbol{\chi}$ based on the governing equations (\eq{eq3} and \eq{eq4}).
\section{Traditional Born Iterative Method}
Traditional Born iterative method is one of the effective algorithms for solving ISPs\cite{slaney1984limitations}.
TBIM reconstructs unknown scatterers with the knowledge of incident and scattered fields via an alternate update process.
TBIM first establishes the following minimization problem:
\begin{equation}
	\min_{ \boldsymbol{\chi}} ||\mathbf{E}^{s} - \mathbf{G}_S \boldsymbol{\chi} \mathbf{E}^{t} ||^2 \,.
	\label{eq5}
\end{equation}
The optimal solution $\boldsymbol{\chi}_{next}$ of \eq{eq5} serves as the update of $\boldsymbol{\chi}$.

Then, the total field $\mathbf{E}^{t}$ can be updated with $\boldsymbol{\chi}_{next}$:
\begin{equation}
	\mathbf{E}^{t}_{next} = (\mathbf{I} - \mathbf{G}_D \boldsymbol{\chi}_{next})^{-1} \mathbf{E}^{i}
	\label{eq6}
\end{equation}
In TBIM, \eq{eq5} and \eq{eq6} are applied alternatively in the iterative process until $\mathbf{E}^{t}$ and $\boldsymbol{\chi}$ satisfy the stop criterion.
In the first update of $\boldsymbol{\chi}$, the incident field $\mathbf{E}^{i}$ is used to approximate the total field $\mathbf{E}^{t}$ in \eq{eq5}. Here, \eq{eq5} is viewed and solved as a least-square problem as stated in \cite{wang1989iterative, chew1990reconstruction}. \eq{eq6} is solved directly by MoM where the computation of $\mathbf{G}_D$ is accelerated by fast Fourier transform.  The algorithm flow of TBIM is summarized in Algorithm \ref{BIMalg}. 

\begin{algorithm}
	\caption{Traditional Born Iterative Method}
	\label{BIMalg}
	\begin{algorithmic}
		\item[] \textbf{Initialization:} $\mathbf{E}_0^{t} =  \mathbf{E}^{i}$, $k=0$, $stopflag = 1$
		\WHILE{$stopflag$}
		\STATE 	\textbf{step 1:} $\boldsymbol{\chi}_{k+1} = \mathop{\arg\min}_{ \boldsymbol{\chi}} ||\mathbf{E}^{s} - \mathbf{G}_S  \boldsymbol{\chi} \mathbf{E}_k^{t} ||^2$
		\STATE \textbf{step 2:} $\mathbf{E}_{k+1}^{t} = 	(\mathbf{I} - \mathbf{G}_D \boldsymbol{\chi}_{k+1})^{-1} \cdot  \mathbf{E}^{i}$
		\STATE \textbf{step 3:} $k=k+1$
		\IF{ stop criteria is satisfied}
		\STATE stopflag = 0
		\ENDIF
		\ENDWHILE
		\item[] \textbf{output:} $\boldsymbol{\chi}_k, \mathbf{E}_{k}^{t}$
	\end{algorithmic}
\end{algorithm}
\section{Neural Born Iterative Method}
Neural Born iterative method is motivated by the physics-informed supervised residual learning.
PhiSRL is built by interpreting ResNet as the fixed-point iterative method\cite{shan2021physics}.
PhiSRL aims to solve the matrix equation $\mathbf{A}\mathrm{x} = b$ and its $k+1$-th update equation can be formulated as:
\begin{equation}
	\mathrm{x}_{k+1} = \mathrm{x}_{k}+\mathcal{F}_{k+1}(\mathbf{R}_{k},\Theta_{k+1})\,,
	\label{eq7}
\end{equation}
where $\mathcal{F}_{k+1}$ and $\Theta_{k+1}$ denote the CNN and corresponding parameter set, $\mathbf{R}_k =  b-\mathbf{A}\mathrm{x}_k$ denotes the residual of $k$-th iteration.
It can be observed from \eq{eq7} that PhiSRL applies CNNs to iteratively update the candidate solutions regarding the calculated residuals, which further motivates the application of PhiSRL to emulate the computing process of TBIM.

NeuralBIM are designed by applying CNNs to alternatively update both total fields and contrasts in TBIM.
The difference between TBIM and NeuralBIM lies in that NeuralBIM applies CNNs to learn the update rules instead of the hand-crafted rules (\eq{eq5} and \eq{eq6}).
Combining \eq{eq5} and \eq{eq7}, the update of $\boldsymbol{\chi}$ can be written as:
\begin{equation}
	\begin{aligned}
		\mathcal{R}^{\mathbf{E}^{s}}_k &= \mathbf{E}^{s} - \mathbf{G}_S \boldsymbol{\chi}_k \mathbf{E}^{t}_k \,, \\
		\boldsymbol{\chi}_{k+1} &= \boldsymbol{\chi}_k + \mathcal{F}_{k+1}^{\chi}(\mathcal{R}^{\mathbf{E}^{s}}_k \oplus \boldsymbol{\chi}_k, \Theta_{k+1}^{\chi}) \,,
	\end{aligned}
	\label{eq8}
\end{equation}
where $	\mathcal{R}^{\mathbf{E}^{s}}_k$ represents the residual of scattered field, $\mathcal{F}_{k+1}^{\chi}$ and $\Theta_{k+1}^{\chi}$ denote the $\boldsymbol{\chi}$-CNN and corresponding parameter set, $\oplus$ is the concatenation of two tensors.
It is worth noting that the input of the $\boldsymbol{\chi}$-CNN adopts the concatenation of the calculated residual and the previous contrast.
This is because $\mathbf{G}_S$ in \eq{eq4} reduces the dimensionality of total fields and results in the loss of information.
Taking $\boldsymbol{\chi}$ as another input of the $\boldsymbol{\chi}$-CNN can enable better performance of inversions according to the authors' trials.
Similarly, the update of total field $\mathbf{E}^{t}$ can be formulated by combining \eq{eq6} and \eq{eq7}:
\begin{equation}
	\begin{aligned}
		\mathcal{R}_{k}^{\mathbf{E}^{i}} &= \mathbf{E}^{i}-(\mathbf{I} - \mathbf{G}_D \boldsymbol{\chi}_{k+1}) \mathbf{E}^{t}_k \,, \\
		\mathbf{E}_{k+1}^{t} &= \mathbf{E}_{k}^{t} + \mathcal{F}_{k+1}^{\mathbf{E}^{t}}(\mathcal{R}_{k}^{\mathbf{E}^{i}}, \Theta_{k+1}^{\mathbf{E}^{t}})  \,,
	\end{aligned}
	\label{eq9}
\end{equation}
where $	\mathcal{R}_{k}^{\mathbf{E}^{i}}$ is the residual of the incident field, $\mathcal{F}_{k+1}^{\mathbf{E}^{t}}$ and $\Theta_{k+1}^{\mathbf{E}^{t}}$ are the $\mathbf{E}^{t}$-CNN and corresponding parameter set.
The algorithm flow of NeuralBIM is summarized in Algorithm \ref{NeuralBIMalg}.
\begin{algorithm}
	\caption{Neural Born Iterative Method}
	\label{NeuralBIMalg}
	\begin{algorithmic}
		\item[] \textbf{initialization:} $\mathbf{E}_0^{tot} =  \mathbf{E}^{inc}$, $\boldsymbol{\chi}_0=0$, $k=0$, $MaxIter$
		\WHILE{$k<MaxIter$}
		\STATE \textbf{step 1:} $\mathcal{R}^{\mathbf{E}^{s}}_k = \mathbf{E}^{s} - \mathbf{G}_S  \boldsymbol{\chi}_k \mathbf{E}^{t}_k$
		\STATE 	\textbf{step 2:} $ \boldsymbol{\chi}_{k+1} = \boldsymbol{\chi}_k + \mathcal{F}_{k+1}^{\chi}(\mathcal{R}^{\mathbf{E}^{s}}_k \oplus \boldsymbol{\chi}_k, \Theta_{k+1}^{\chi})$
		\STATE \textbf{step 3:} $\mathcal{R}_{k}^{\mathbf{E}^i} = \mathbf{E}^{i}-(\mathbf{I} - \mathbf{G}_D \boldsymbol{\chi}_{k+1}) \mathbf{E}^{t}_k$ 
		\STATE \textbf{step 4:} $\mathbf{E}_{k+1}^{t} = \mathbf{E}_{k}^{t} + \mathcal{F}_{k+1}^{E^{t}}(\mathcal{R}_{k}^{\mathbf{E}^i}, \Theta_{k+1}^{\mathbf{E}^{t}})$
		\STATE \textbf{step 5:} $k=k+1$
		\ENDWHILE
		\item[] \textbf{output:} $\boldsymbol{\chi}_{MaxIter}$, $\mathbf{E}_{MaxIter}^{t}$
	\end{algorithmic}
\end{algorithm}

In this paper, NeuralBIM is assumed to have 7 iterative modules as depicted in \fig{BIMNN}.
\fig{BIMNNblock} illustrates the detailed structure of the iterative block.
The $\boldsymbol{\chi}$-CNN and $\mathbf{E}^{t}$-CNN to modify contrasts and total fields share the same structure but different parameter sets.
The employed CNN comprises five stacked operations including $3 \times 3$ convolution, batch normalization and Tanh nonlinearity, as shown in \fig{BIMNNblock}.
Tanh nonlinearity can provide both positive and negative values, which allows an adaptive modification of the candidate solution.
The input and output channels are also denoted in the \fig{BIMNNblock}.
The $c_1$ and $c_2$ are $4$ and $2$ for updating lossy contrasts, and $2$ and $2$ for updating the total fields.

The training schemes of NeuralBIM include the supervised and unsupervised learning schemes:
\par
\textbf{Supervised learning scheme:}
The supervised learning scheme trains the NeuralBIM with the total fields and contrasts as labels.
The objective function is defined as:
\begin{equation}
	obj_{su} = loss_{\mathbf{E}^{t}} + loss_{\boldsymbol{\chi}}\,, 
	\label{eq10}
\end{equation}
where $loss_{\mathbf{E}^{t}}$ and $loss_{\boldsymbol{\chi}}$ denote the losses of total fields and contrasts respectively.
The $loss_{\mathbf{E}^{t}}$ is defined as the mean squared error (MSE):
\begin{equation}
	loss_{\mathbf{E}^{t}} = 	\frac{1}{N_{{\mathbf{E}^{t}}^{\prime}}}||{\mathbf{E}^{t}}^{*} - {\mathbf{E}^{t}}^{\prime}||_{F}^2 \,,
	\label{eq11}
\end{equation}
where ${\mathbf{E}^{t}}^{*}$ and ${\mathbf{E}^{t}}^{\prime}$ denote the ground-truth and predicted total field, $N_{{\mathbf{E}^{t}}^{\prime}}$ denotes the element number in the vector of ${\mathbf{E}^{t}}^{\prime}$, $||\cdot||_F$ represents Frobenius norm.
The $loss_{\boldsymbol{\chi}}$ adds the TV regularization item to the MSE of contrasts:
\begin{equation}
	loss_{\boldsymbol{\chi}} = \frac{1}{N_{\boldsymbol{\chi}^{\prime}}}||\boldsymbol{\chi}^{*} - \boldsymbol{\chi}^{\prime}||_{F}^2 + \alpha\frac{1}{N_{\boldsymbol{\chi}^{\prime}}}||\nabla \boldsymbol{\chi}^{\prime}||_{1} \,,
	\label{eq12}
\end{equation}
where $\boldsymbol{\chi}^{*}$ and $\boldsymbol{\chi}^{\prime}$ are the ground-truth and inverted contrast, $N_{\boldsymbol{\chi}^{\prime}}$ denotes the element number in the vector of $\boldsymbol{\chi}^{\prime}$, $||\cdot||_1$ is L1 norm, $\alpha$ is fixed as $0.0001$.
In the supervised training scheme, NeuralBIM is trained to generate accurate predictions of total fields and contrasts It takes a certain amount of time to generate training data samples by applying MoM to solve \eq{eq3} and \eq{eq4}. This scheme is also different from the practical applications where total fields and contrasts are unknown.
\par
\textbf{Unsupervised learning scheme:}
In the unsupervised learning scheme, total fields and contrasts are unknown when training NeuralBIM.
The training of NeuralBIM is constrained and  guided by the governing equations of ISPs as formulated in \eq{eq3} and \eq{eq4}.
The objective function is defined as:
\begin{equation}
	obj_{unsu} = res_{\mathbf{E}^{i}} + res_{\mathbf{E}^{s}}\,, 
	\label{eq13}
\end{equation}
The $res_{\mathbf{E}^{i}}$ is formulated based on \eq{eq3}:
\begin{equation}
	res_{\mathbf{E}^{i}} = \frac{1}{N_{{\mathbf{E}^{i}}}}||(\mathbf{I} - \mathbf{G}_D \boldsymbol{\chi}^{\prime}) {\mathbf{E}^{t}}^{\prime} - \mathbf{E}^{i}||_{F}^2 \,.
	\label{eq14}
\end{equation}
The $res_{\mathbf{E}^{s}}$ is defined based on \eq{eq4}:
\begin{equation}
	res_{\mathbf{E}^{s}} = \frac{1}{N_{\mathbf{E}^{s}}}|| \mathbf{E}^{s} - \mathbf{G}_S {\boldsymbol{\chi}}^{\prime} {\mathbf{E}^{t}}^{\prime}||_{F}^2 + \alpha\frac{1}{N_{\boldsymbol{\chi}^{\prime}}}||\nabla \boldsymbol{\chi}^{\prime}||_{1} \,,
	\label{eq15}
\end{equation}
where $N_{{\mathbf{E}^{i}}}$ and $N_{\mathbf{E}^{s}}$ denote the element number in the vector of $\mathbf{E}^{i}$ and $\mathbf{E}^{s}$, $||\nabla \boldsymbol{\chi}||_{1}$ is the TV regularization and $\alpha$ is fixed as $0.0001$.
It is noted that the incident and scattered fields in \eq{eq14} and \eq{eq15} are assumed to be known in ISPs and the Green functions are determined by the measurement setup of ISPs. 
The unsupervised training scheme aims to train NeuralBIM to generate solutions simultaneously satisfying \eq{eq3} and \eq{eq4} by leveraging the existing physical law and information. 
The TV regularization can stabilize the training process by enforcing the boundaries of reconstructions.
\par
The computational complexity of NeuralBIM is determined by two parts: the matrix multiplication in calculating residuals and the basic computation in the CNN.
In a single iteration, NeuralBIM has two subsequent branches to update $\boldsymbol{\chi}$ and $\mathbf{E}^{t}$ respectively.
The numbers of transmitters, receivers and pixels are denoted as $T$, $R$ and $P$ for simplification. 
In the $\boldsymbol{\chi}$-branch, the complexity of calculating $\mathcal{R}^{\mathbf{E}^{s}}$ is $O(TRP)$ due to the dense matrix multiplication.
The complexity of the $\boldsymbol{\chi}$-CNN is dominated by convolution operations. For a single convolutional layer, the complexity is $O(PK^{w}K^{h}C_{\chi}^{i}C_{\chi}^{o})$, where $K^{w}$ and $K^{h}$ are the width and height of the convolutional kernel, $C_{\chi}^{i}$ and $C_{\chi}^{o}$ are the input and output channels.
Assuming the $\boldsymbol{\chi}$-CNN has $L$ layers, its complexity can be written as: $O(\sum_{l=1}^{L}PK^{w}_lK^{h}_lC^{i}_{\chi, l}C^{o}_{\chi, l})$. 
Note that all feature maps output by the convolutional layers contain $P$ pixels.
Similarly, in the $\mathbf{E}^{t}$-branch, the complexity of the calculation of $\mathcal{R}_{k}^{E^i}$ and the $\mathbf{E}^{t}$-CNN are $O(TP^2)$ and $O(\sum_{l=1}^{L}PK^{w}_lK^{h}_lC_{E^{t}, l}^{i}C_{E^{t}, l}^{o})$ respectively.
The complexity of calculating $\mathcal{R}_{k}^{E^i}$ can be further reduced by fast Fourier transform in the case of large-scale problems.
For NeauralBIM with $N$ iterations, the total computational complexity is $O(NTRP+N\sum_{l=1}^{L}PK^{w}_lK^{h}_lC^{i}_{\chi, l}C^{o}_{\chi, l} + NTP^2 + N\sum_{l=1}^{L}PK^{w}_lK^{h}_lC_{E^{t}, l}^{i}C_{E^{t}, l}^{o})$.
\section{Numerical Results}
\begin{figure}
	\centering
	\includegraphics[width=0.9\linewidth]{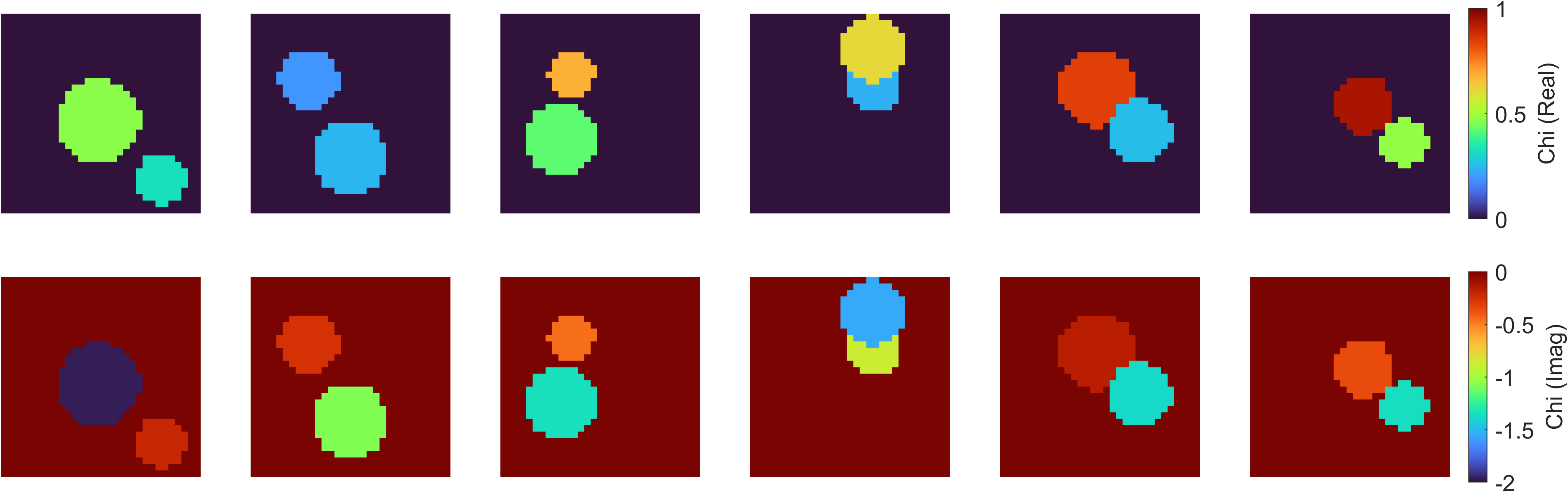}
	\caption{Examples of scatterers in the synthetic data set. Two cylinders have random locations and radii, as  detailed in \tab{trainTab}.}
	\label{chiexample}
\end{figure}
\begin{figure}
	\centering
	\includegraphics[width=0.8\linewidth]{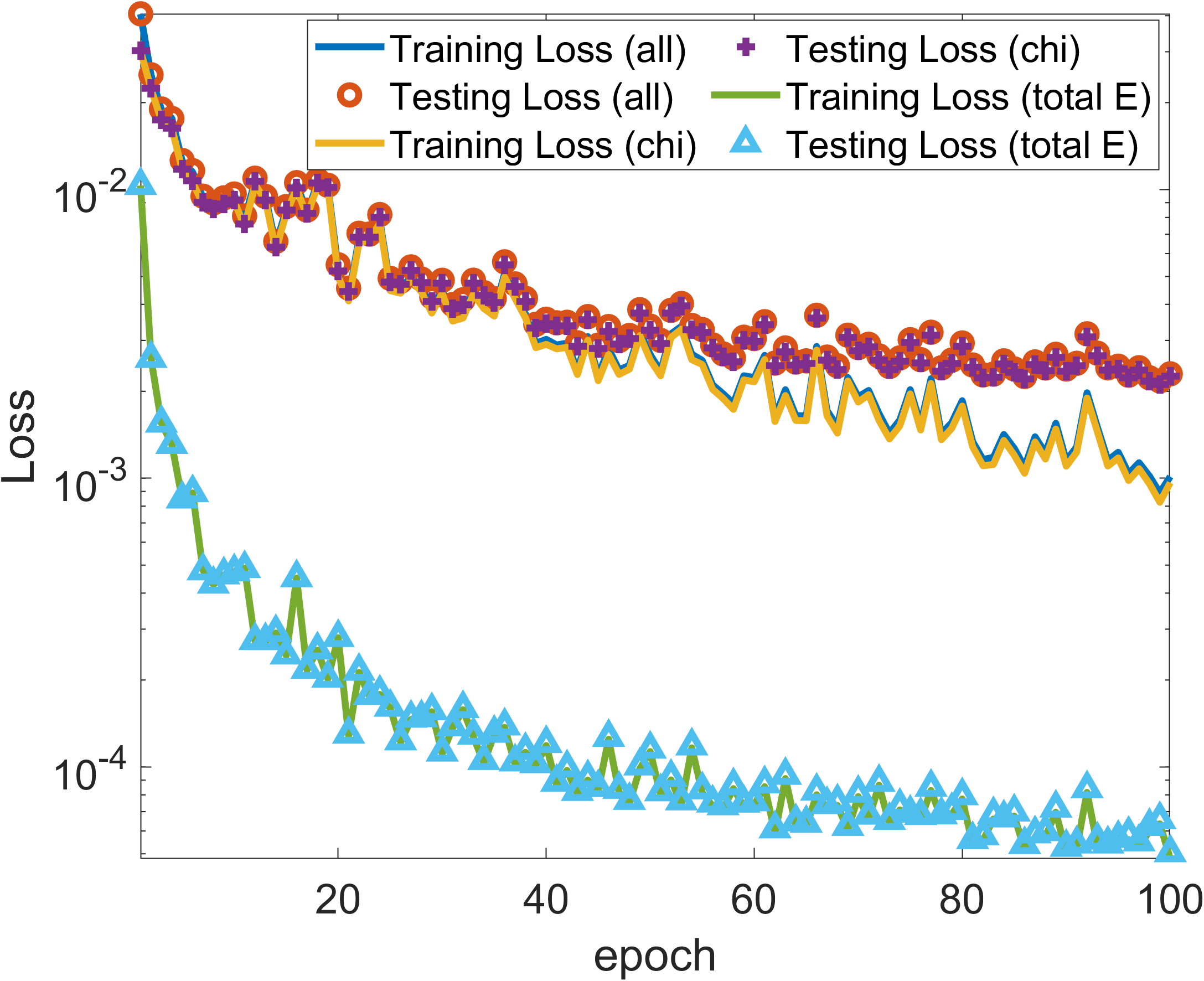}
	\caption{Convergence curve of supervised NeuralBIM. Training and testing loss (all, chi, total E) denote $obj_{su}$, $loss_{\boldsymbol{\chi}}$, and $loss_{\mathbf{E}^{t}}$.}
	\label{suloss}
\end{figure}
\begin{figure}
	\centering
	\subfigure[ ]{
		\centering
		\includegraphics[width=1\linewidth]{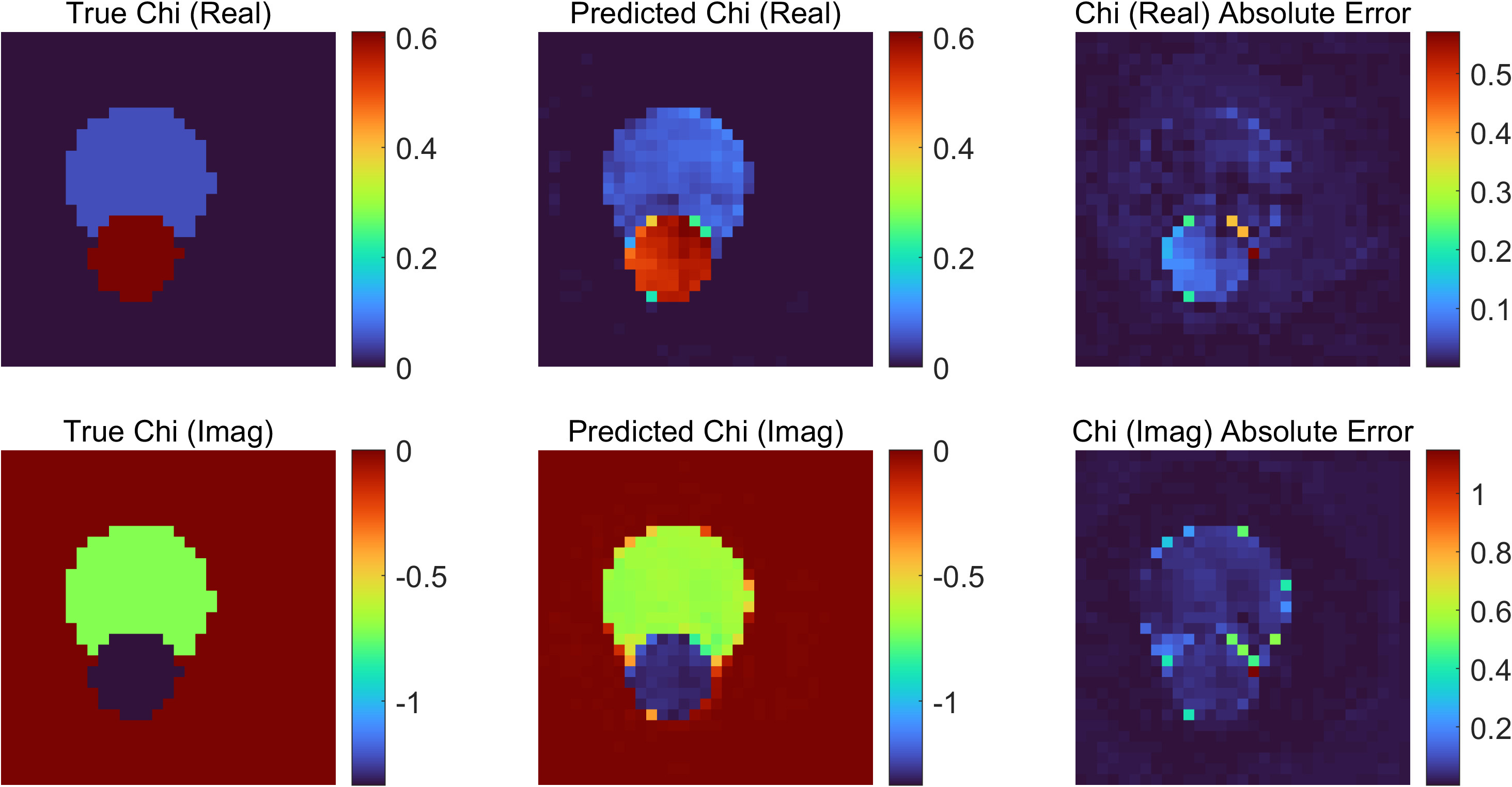}
	}
	\subfigure[ ]{
		\centering
		\includegraphics[width=1\linewidth]{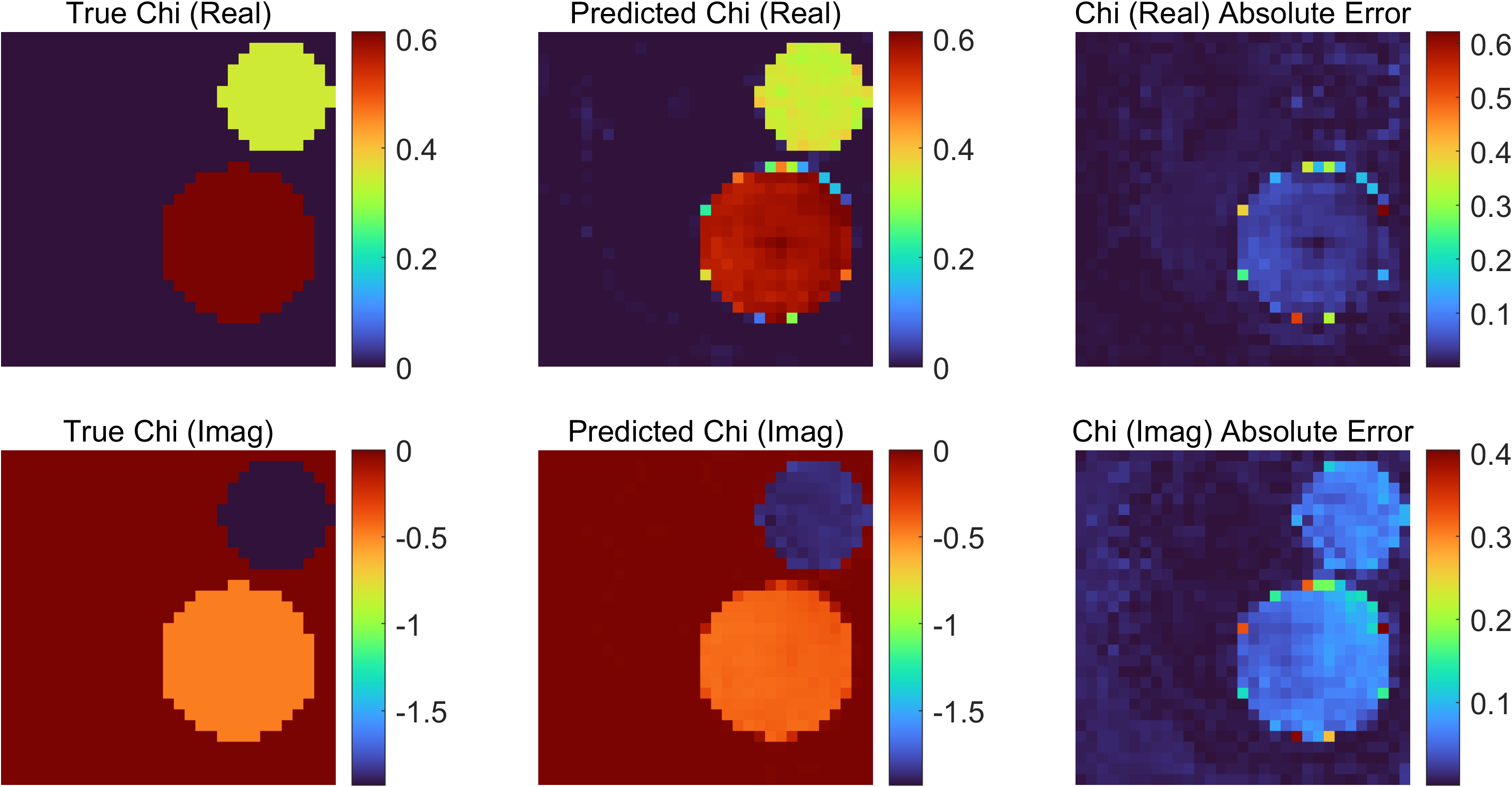}
	}
	\caption{Comparison between ground truth and scatterers inverted by supervised NeuralBIM. From left to right are: ground truth, reconstruction and their absolute error distribution.}
	\label{suchi}
\end{figure}
In this section, we first validate both supervised and unsupervised NeuralBIM with synthetic data inversion.
The efficacy of unsupervised NeuralBIM is further verified with experimental data inversion.
Supervised and unsupervised NeuralBIM are implemented in Pytorch and computed on three Nvidia V100 GPUs.
The Adam optimizer is adopted to train NeuralBIM. The learning rate is initialized as $0.002$ and decayed by $\sim \times 0.8$ every 20 epochs.
\begin{table}[h]
	\centering
	\caption{Training Settings in Synthetic Data Inversion}
	\label{trainTab}
	\begin{tabular}{cccc}
		\hline
		Cylinder & Radius & Contrast (Real) & Contrast (Imag)\\ \hline
		A & 0.015-0.035$m$ & [0, 1] & [-1, 0] \\
		B & 0.015-0.035$m$ & [0, 1] & [-2, -1] \\ \hline
	\end{tabular}
\end{table}
\subsection{Synthetic Data Inversion}
As depicted in \fig{modelsetup}, $D$ has a size of $0.15m \times 0.15m$ and it is uniformly discretized into $32 \times 32$ grids. 
32 transmitters and receivers are evenly distributed on a circle outside $D$ of which radius is $1. 67m$. 
The incident frequency is 3GHz. 
Two cylinders are randomly located inside $D$ as summarized in \tab{trainTab}.
\fig{chiexample} shows several examples of scatterer shapes.
The data set is generated by MoM. 
Supervised and unsupervised NeuralBIM are trained and tested on the same data set.
The data set comprises 20000 data samples of which 80\% are for training and 20\% for testing.
The initial guess of the total field adopts the incident field.
BP method is applied to generate the initial guess of the contrast. 
\subsubsection{Supervised Neural Born Iterative Method}
\begin{figure}
	\centering
	\subfigure[ ]{
		\centering
		\includegraphics[width=1\linewidth]{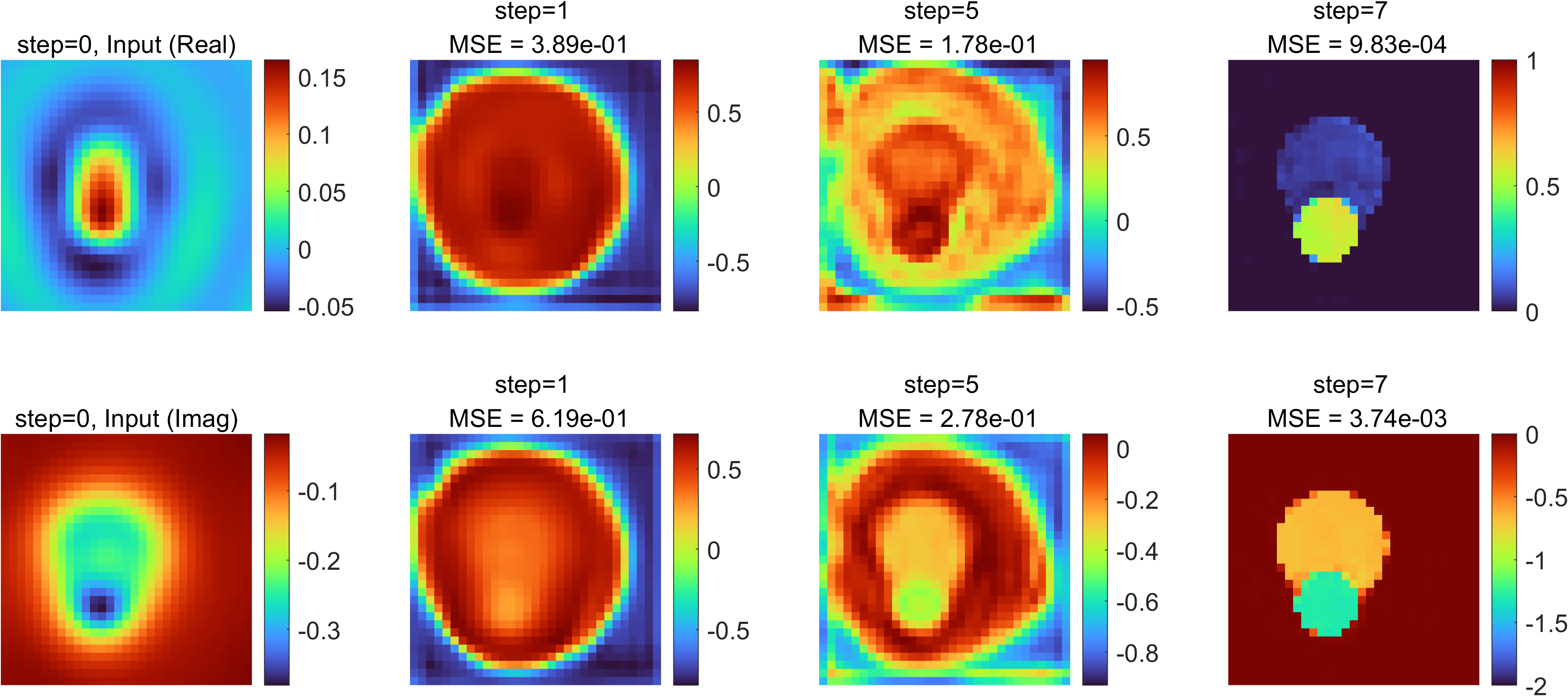}
	}
	\subfigure[ ]{
		\centering
		\includegraphics[width=1\linewidth]{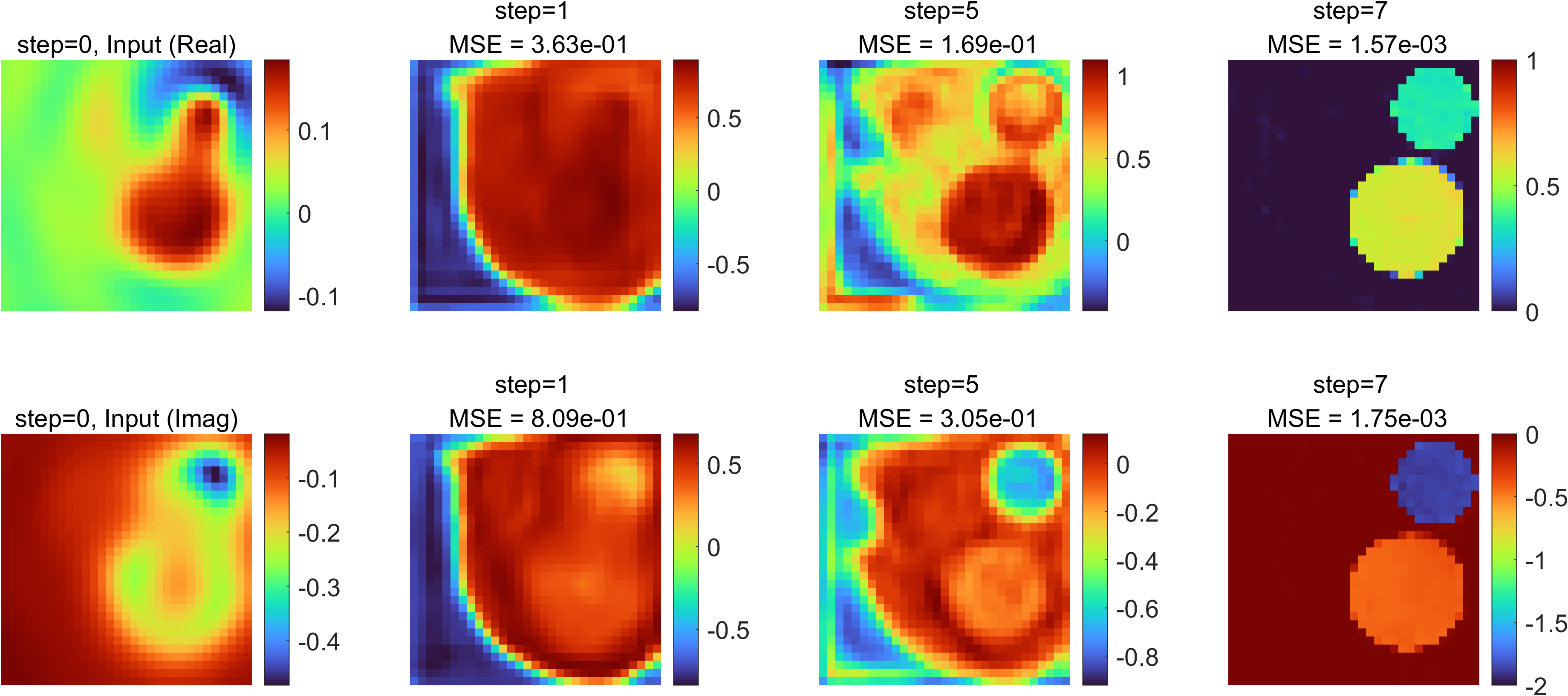}
	}
	\caption{Updated reconstructions of contrasts in the iterative computing of supervised NeuralBIM. From left to right are: initial guess, reconstructions at the first, fifth and seventh iteration. In each sub-panel, the first and second row show the real and imaginary parts.}
	\label{sutotall}
\end{figure}
\begin{figure}
	\subfigure[]{
	\centering
	\includegraphics[width=0.95\linewidth]{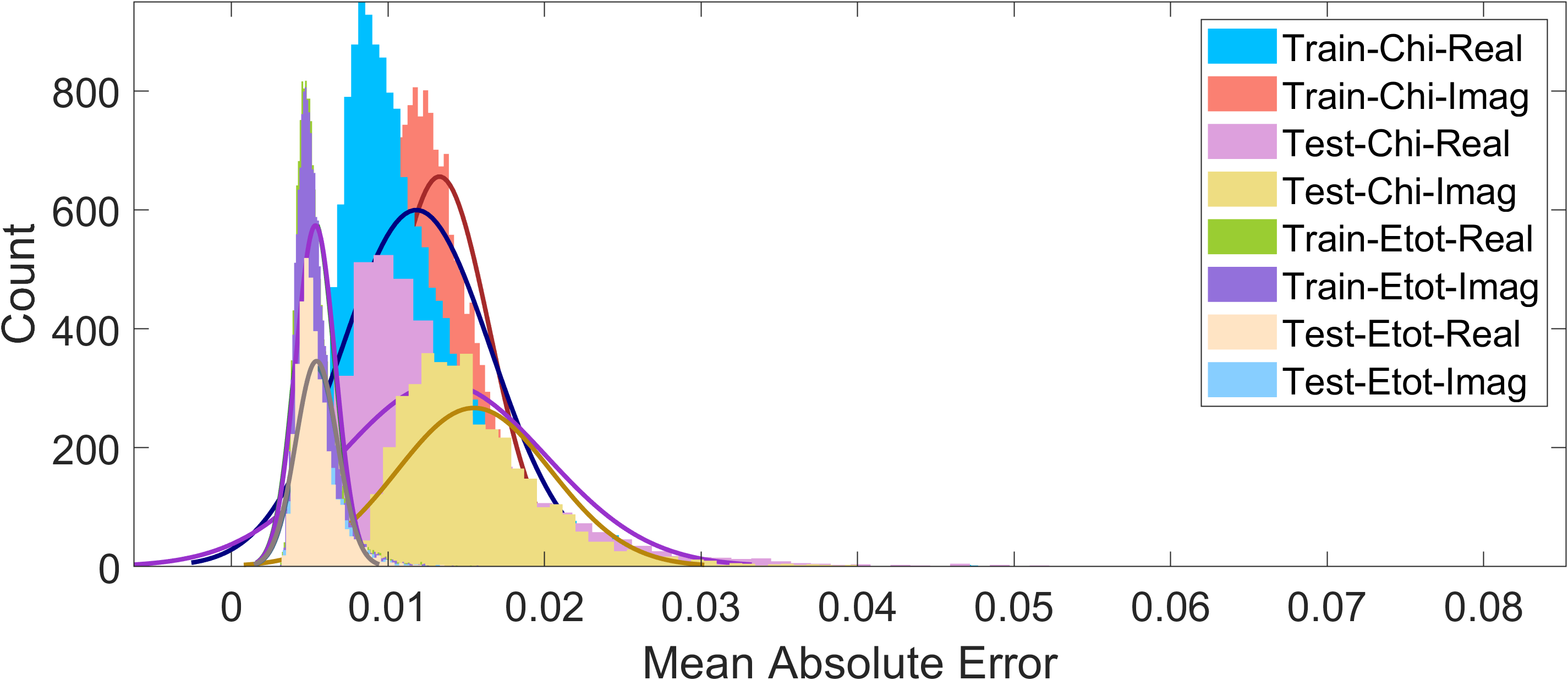} 
    \label{suMAEall}}
	\subfigure[]{
	\centering
	\includegraphics[width=0.95\linewidth]{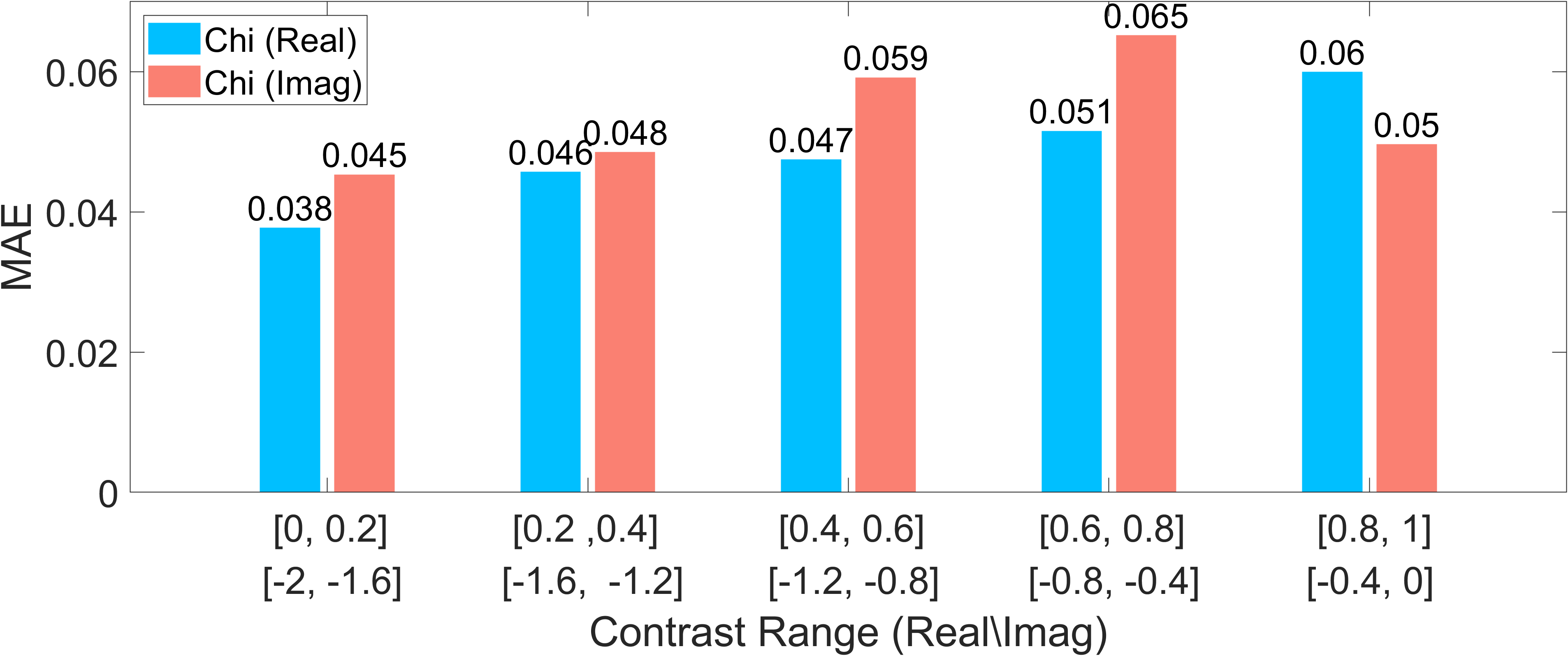}
	\label{suMAEchi}}
	\caption{MAE histograms of contrasts and total fields inverted by supervised NeuralBIM. (a) is a histogram of the reconstruction MAE of each data sample in training and testing data sets. (b) is a histogram of the reconstruction MAE of each cylinder at different contrast ranges for all data samples. Chi, Etot, Real, Imag denote the contrast, total field, real part and imaginary part in the legend.}
\end{figure}

\fig{suloss} plots the convergence curves of $obj_{su}$, $loss_{\mathbf{E}^{t}}$ and $loss_{\boldsymbol{\chi}}$. 
All of them drop steadily as the training progresses. 
It can be observed that $loss_{\boldsymbol{\chi}}$ dominates $loss_{\mathbf{E}^{t}}$ with the small value of $loss_{\mathbf{E}^{t}}$.
The testing $loss_{\boldsymbol{\chi}}$ is a little higher than the training one which leads to overfitting.
The overfitting can be alleviated by introducing better regularizations or stopping training earlier.

{\fig{suchi} depicts the comparisons between ground truth and reconstructions that are randomly chosen in the testing data set. The inverted contrasts are in a good agreement with the ground truth as reflected in the absolute error distributions.
The differences mostly lie on the boundaries of scatterer shapes.

The updated reconstructions of contrasts are illustrated in \fig{sutotall}.
The update of contrast begins with the initial guess generated by BP method and it improves gradually with the increase of iterations.}

The mean absolute error (MAE) histograms of all data samples are charted in \fig{suMAEall}.
The specific means and standard deviations (stds) are summarized in \tab{sutab}.
The MAE histogram of total fields has small means and stds, and this indicates the good reconstruction precisions.
The MAE histogram of testing contrasts has higher mean and std than the one of training contrasts, which is consistent with the convergence curve in the \fig{suloss}.
\fig{suMAEchi} charts a histogram of the reconstruction MAE of each cylinder at different contrast ranges for all data samples. The reconstruction MAE is only evaluated for the cylinders and the background is not considered. 
The real and imaginary part ranges of the contrast is evenly divided into five intervals respectively.
For real parts, the five intervals are [0, 0.2], [0.2 ,0.4], [0.4, 0.6], [0.6, 0.8], [0.8, 1.0], and they are [-2, -1.6], [-1.6, -1.2], [-1.2, -0.8], [-0.8, -0.4], [-0.4, 0] for imaginary parts.
Supervised NeuralBIM shows a stable performance regarding a wide range of contrast values.

\begin{figure}
	\centering
	\includegraphics[width=1\linewidth]{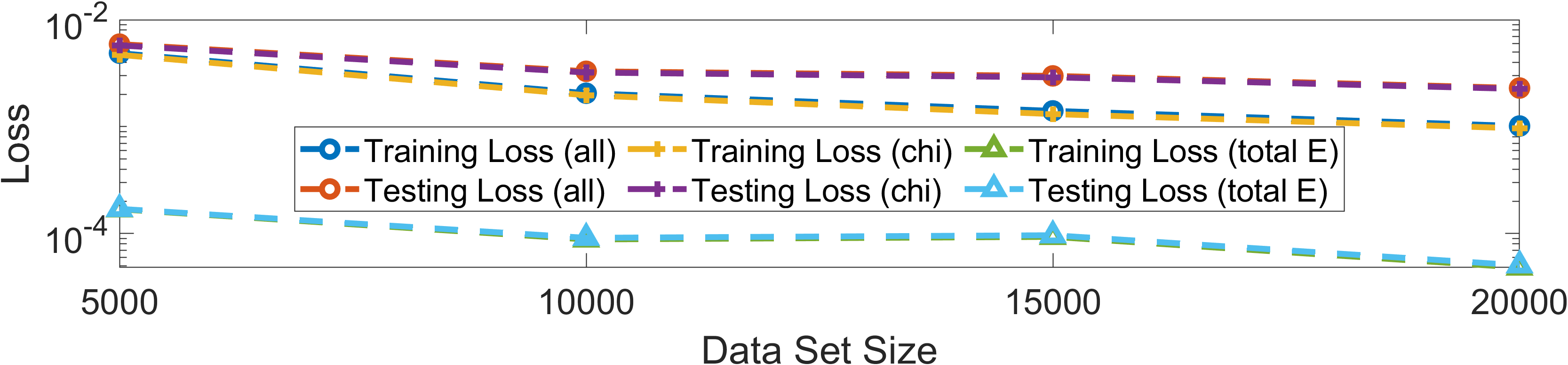}
	\caption{The final losses of  supervised NeuralBIM trained with data sets of different sizes. Training and testing loss (all, chi, total E) denote $obj_{su}$, $loss_{\boldsymbol{\chi}}$, and $loss_{\mathbf{E}^{t}}$.}
	\label{datasetloss}
\end{figure}

The impact of data set sizes on the performance of supervised NeuralBIM is first investigated. The data sets are assumed to have $5000$, $10000$, $15000$, $20000$ samples respectively, of which $80\%$ and $20\%$ are for training and testing. For all data sets, all training processes are assumed to stop at the 100th epoch.
\fig{datasetloss} shows the final losses of  supervised NeuralBIM trained with data sets of different sizes. 
It can be observed that the larger the training data set, the better the reconstruction performance.
When the size of data set increases from $15000$ to $20000$, the testing $loss_{\boldsymbol{\chi}}$ is slightly reduced although the training $loss_{\boldsymbol{\chi}}$ has a noticeable drop. 
Therefore, considering the training time, the data set of $20000$ samples is adopted to train NeuralBIM in this paper.

\begin{table}
	\renewcommand{\arraystretch}{1.3}
	\caption{MAE means and standard deviations of supervised NeuralBIM}
	\label{sutab}
	\centering
	\begin{threeparttable}
		\begin{tabular}{ccc}
			\toprule
			Item \,\, \, & MAE-R\tnote{*} \, (mean/std) \,\,  & MAE-I\tnote{+} \, (mean/std) \,\,\,\,\, \\
			\midrule
			Contrast (train)\tnote{1} & $0.0118/0.0048$ & $0.0133/0.0032$\\
			Contrast (test)\tnote{2} & $0.0135/0.0066$ & $0.0155/0.0049$ \\
			Total field (train)\tnote{3} & $0.0054/0.0013$ & $0.0053/0.0012$ \\
			Total field (test)\tnote{4} & $0.0055/0.0013$ & $0.0054/0.0013$ \\
			\bottomrule
		\end{tabular}
		\begin{tablenotes}
			\setlength{\multicolsep}{0cm}
			\begin{multicols}{2}
				\item[1] contrasts in training data set
				\item[2] contrasts in testing data set
				\item[3] total fields in training data set
				\item[4] total fields in testing data set \\
				\item[*] MAE of real part 	
				\item[+] MAE of imaginary part 
			\end{multicols}
		\end{tablenotes}
	\end{threeparttable}
\end{table}
\begin{figure}
	\centering
	\includegraphics[width=1\linewidth]{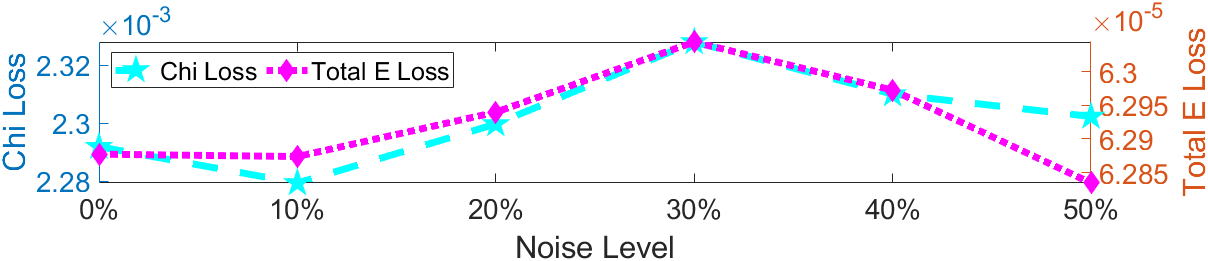}
	\caption{Losses of contrasts and total fields reconstructed by supervised NeuralBIM under different noise levels. Chi and total E denote contrast and total field.}
	\label{noiseloss}
\end{figure}
\begin{figure}
	\centering
	\includegraphics[width=1\linewidth]{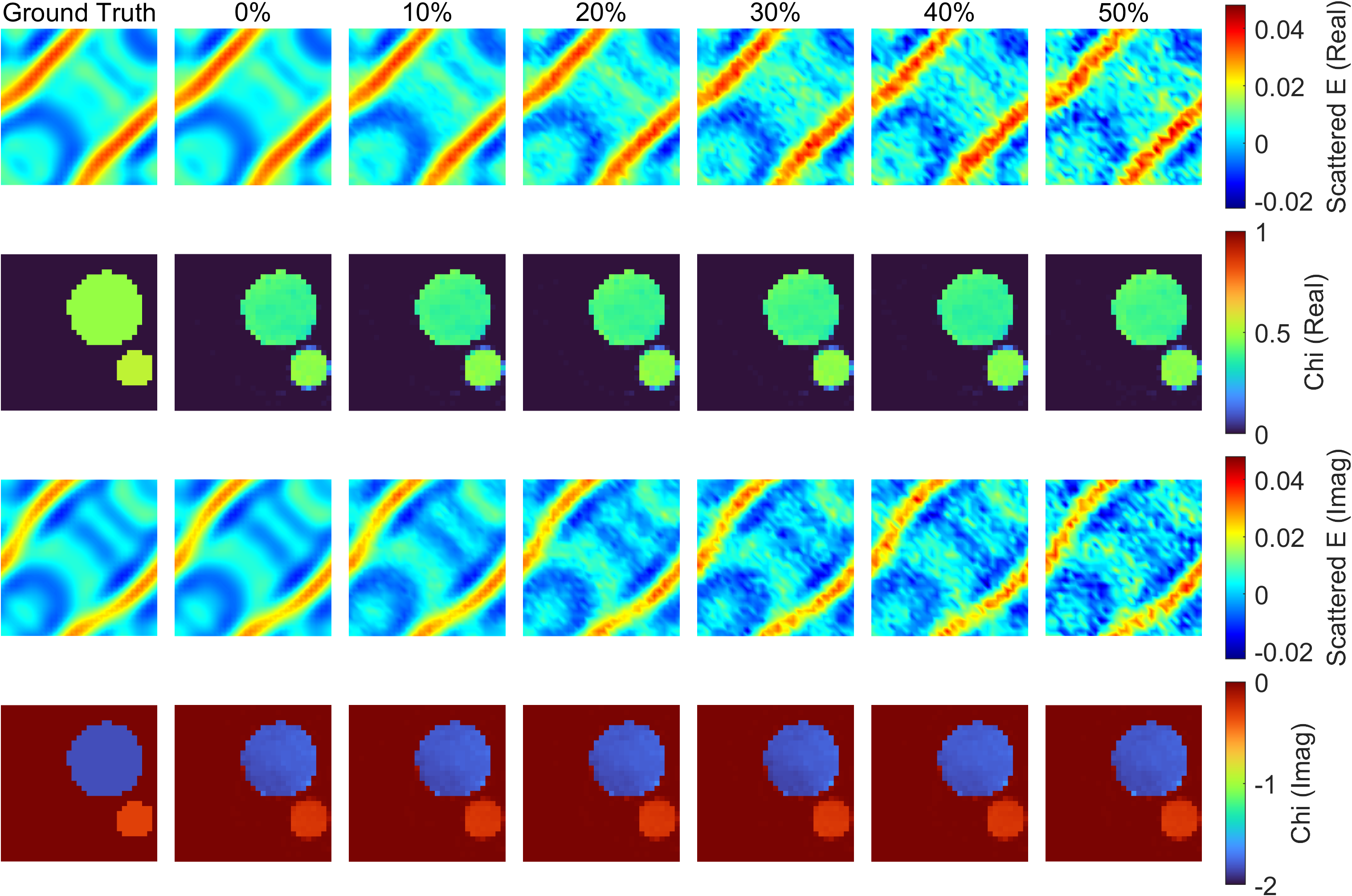}
	\caption{Reconstructed scatterers of supervised NeuralBIM under different noise levels. From left to right are ground truth, reconstructions at the noise level of 0\%, 10\%, 20\%, 30\%, 40\%, 50\%.}
	\label{noisetest}
\end{figure}
\begin{figure}
	\centering
	\includegraphics[width=1\linewidth]{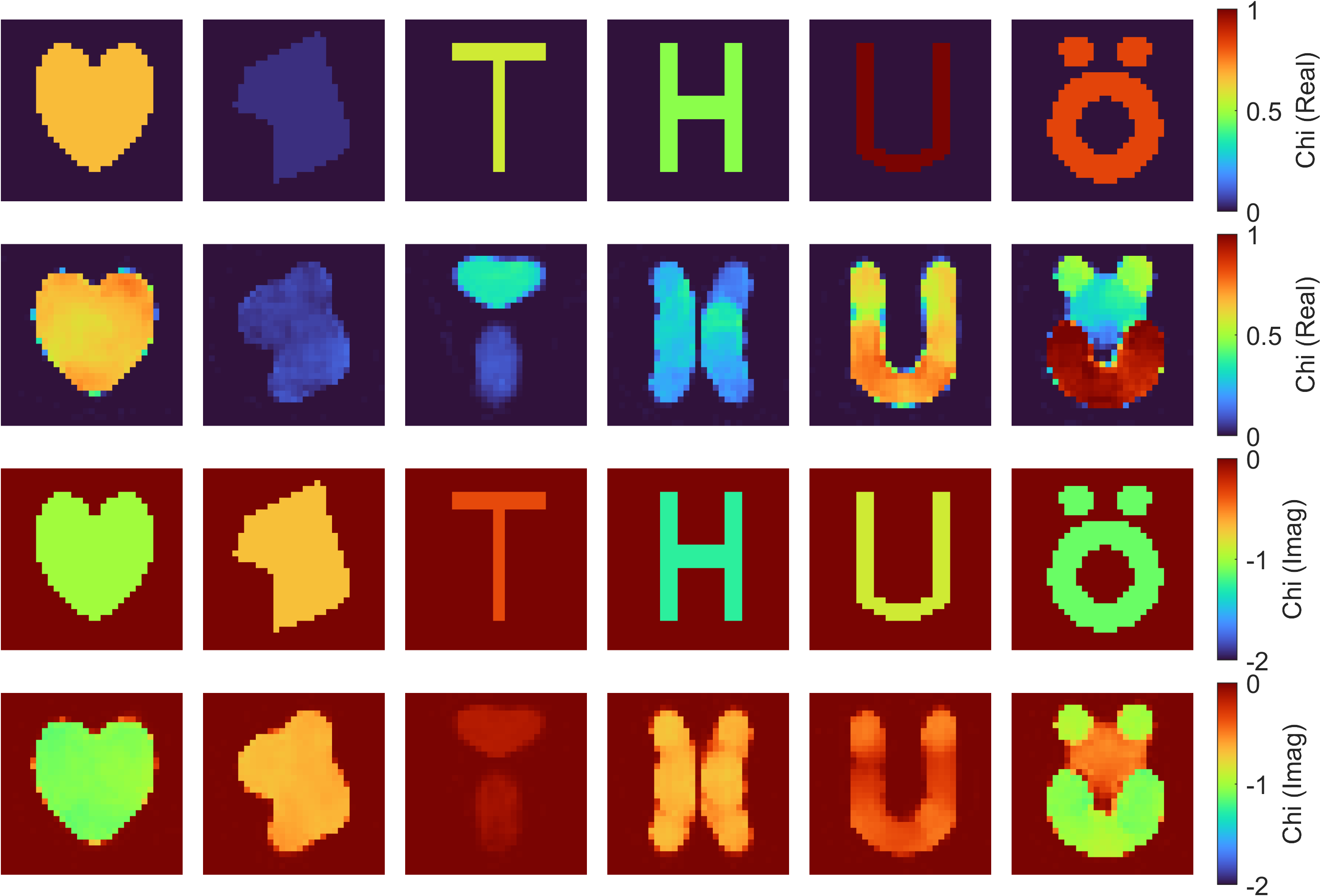}
	\caption{Verification of generalization ability of supervised NeuralBIM on different scatterer shapes.}
	\label{geometrytest}
\end{figure}
\begin{figure}
	\centering
	\includegraphics[width=1\linewidth]{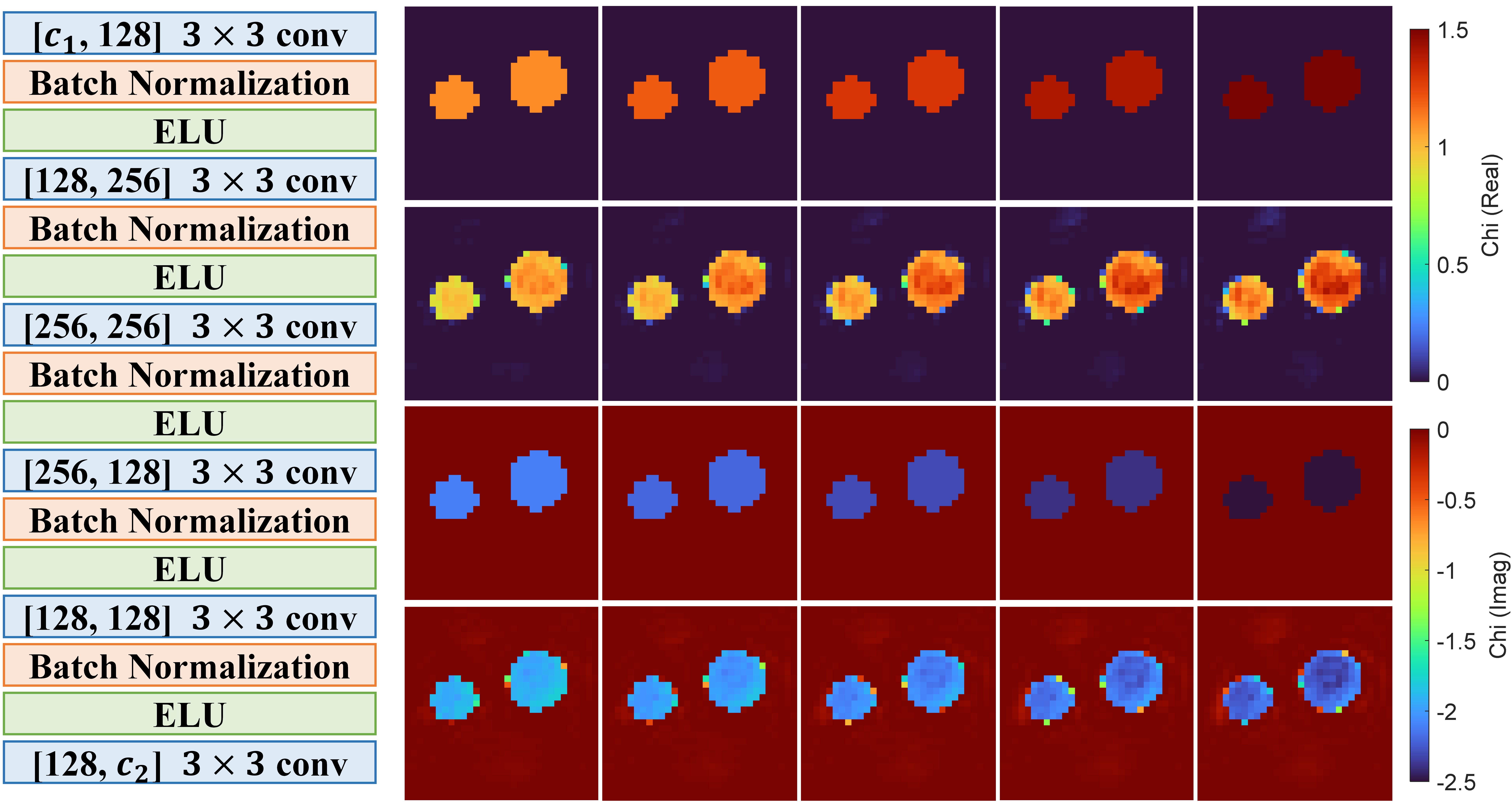}
	\caption{The CNN structure and the reconstructed scatterers in different contrasts. The CNN structure in the re-trained supervised NeuralBIM (left sub-panel) and the reconstructed scatterers in different contrasts (right sub-panel). From left to right, the contrasts are: $1.1-j2.1$, $1.2-j2.2$, $1.3-j2.3$, $1.4-j2.4$, $1.5-j2.5$. }
	\label{highchiresult}
\end{figure}
\begin{figure}
	\centering
	\subfigure[ ]{
		\centering
		\includegraphics[width=0.46\linewidth]{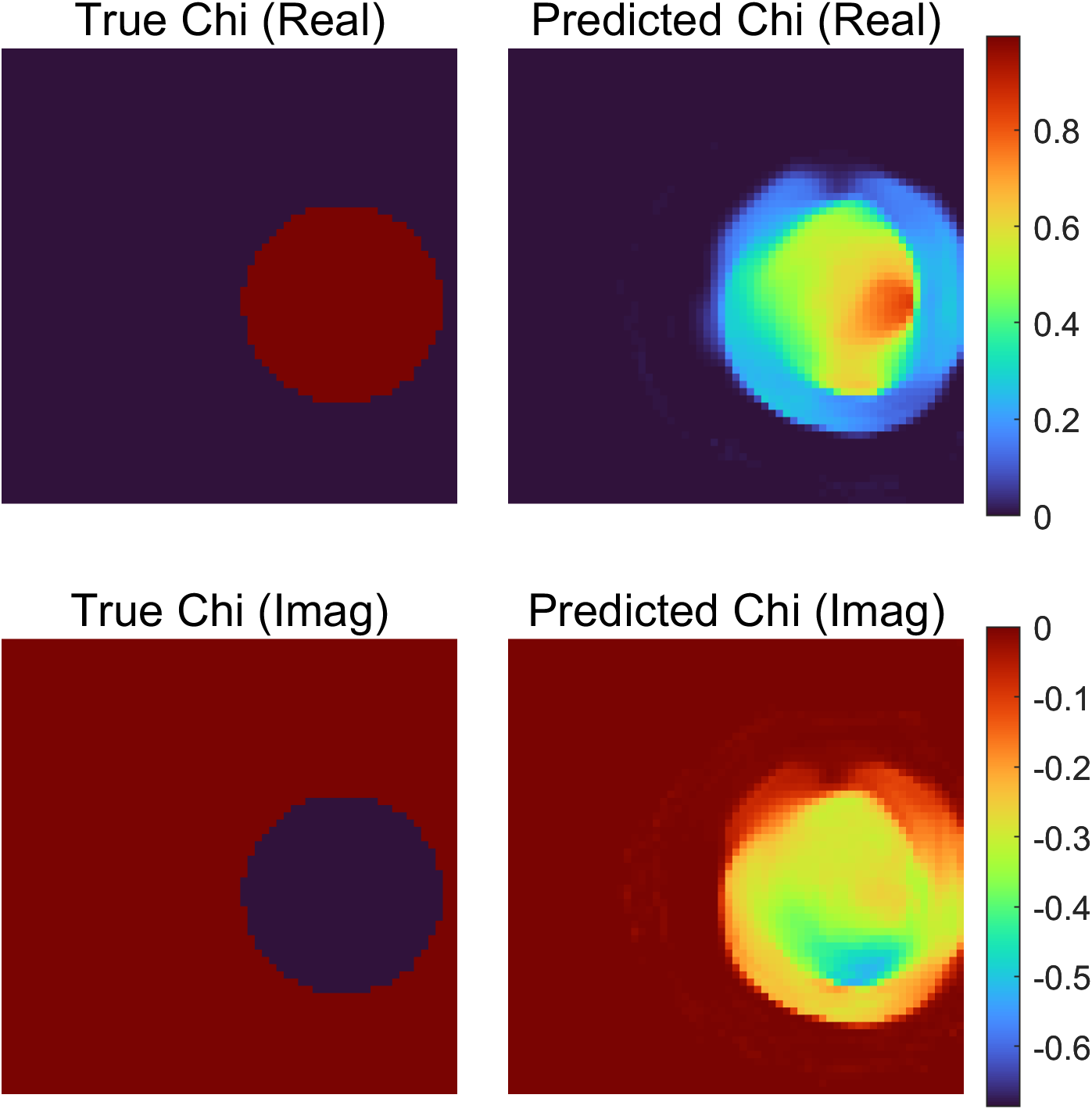}
	}
	\subfigure[ ]{
		\centering
		\includegraphics[width=0.46\linewidth]{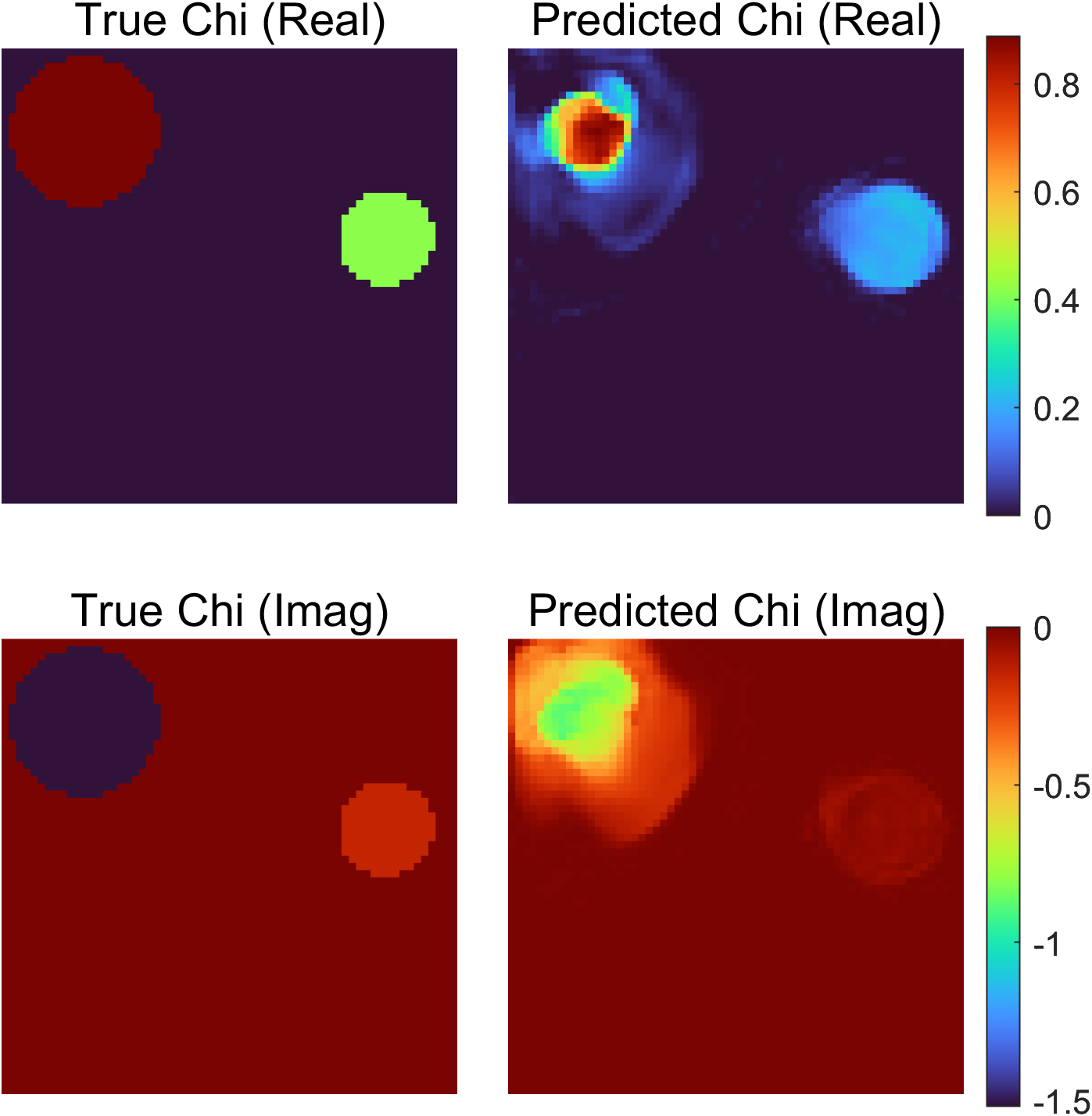}
	}
	\caption{Reconstructions of $64\times 64$ grids. (a): one cylinder, (b): two cylinders. In each sub-panel, from left to right are: ground truth and the reconstructed scatterers.}
	\label{scaleresult}
\end{figure}
\par
The anti-noise capability is then validated.
We randomly select 20 testing data samples and add different levels of Gaussian white noise.
\fig{noiseloss} plots $loss_{\mathbf{E}^{t}}$ and $loss_{\boldsymbol{\chi}}$ of each noise level. 
Both $loss_{\mathbf{E}^{t}}$ and $loss_{\boldsymbol{\chi}}$ fluctuate within a very small range, which demonstrates the strong anti-noise capability of supervised NeuralBIM.
The inverted contrasts are detailed in \fig{noisetest}.
The reconstruction maintains a good and stable quality even at the noise level of 50\%.

The generalization ability is also verified on different scatterer geometries unseen at the training time.
Six different kinds of scatterer geometries are considered as shown in \fig{geometrytest}.
The comparisons between the ground truth and reconstructions are demonstrated in \fig{geometrytest}.
Supervised NeuralBIM can perform reliable inversion although the scatterer geometries are not included in the training data set.
The discrepancy primarily lies in the tiny structures that are hard to distinguish. 

Furthermore, supervised NeuralBIM is employed to invert the scatterers with higher contrasts. Supervised NeuralBIM with CNN structures as shown in \fig{BIMNNblock} is insensitive to the higher contrasts, which could be caused by that Tanh functions are saturating nonlinearities. The modified CNN model with ELU nonlinearities \cite{clevert2015fast} is adopted as shown in \fig{highchiresult}. The modified supervised NeuralBIM is re-trained under the same training settings. It demonstrates an improved performance on inverting scatterers on higher contrasts, as shown in \fig{highchiresult}.

Supervised NeuralBIM is also tested whether it can be generalized to solve large-scale problems when trained for small-scale problems.
Supervised NeuralBIM is trained on $32 \times 32$ grids.
Then, it is applied to perform inversion of $64\times64$ grids.
According to \eq{eq8}, in the $k$-th iteration, $\boldsymbol{\chi}$-CNN is dependent on the concatenation of $\mathcal{R}^{\mathbf{E}^{s}}_k$ and $\boldsymbol{\chi}_k$.
Note that $\mathcal{R}^{\mathbf{E}^{s}}_k$ and $\boldsymbol{\chi}_k$ have the same size and are concatenated in the channel dimension.
The size of  $\mathcal{R}^{\mathbf{E}^{s}}_k$ is determined by the number of transmitters and recievers.
When the size of $\boldsymbol{\chi}_k$ is $64\times64$, the number of transmitters and recievers need be set as $64$.
Such change of the measurement configuration results in a reduced reconstruction performance of Supervised NeuralBIM.   
\fig{scaleresult} demonstrates two different reconstructed scatterers of $64\times64$ grids.
The inverted results can reflect the locations and shapes of scatterers but the corresponding contrast values are not accurate enough.
\subsubsection{Unsupervised Neural Born Iterative Method}
\begin{figure}
	\centering
	\includegraphics[width=0.8\linewidth]{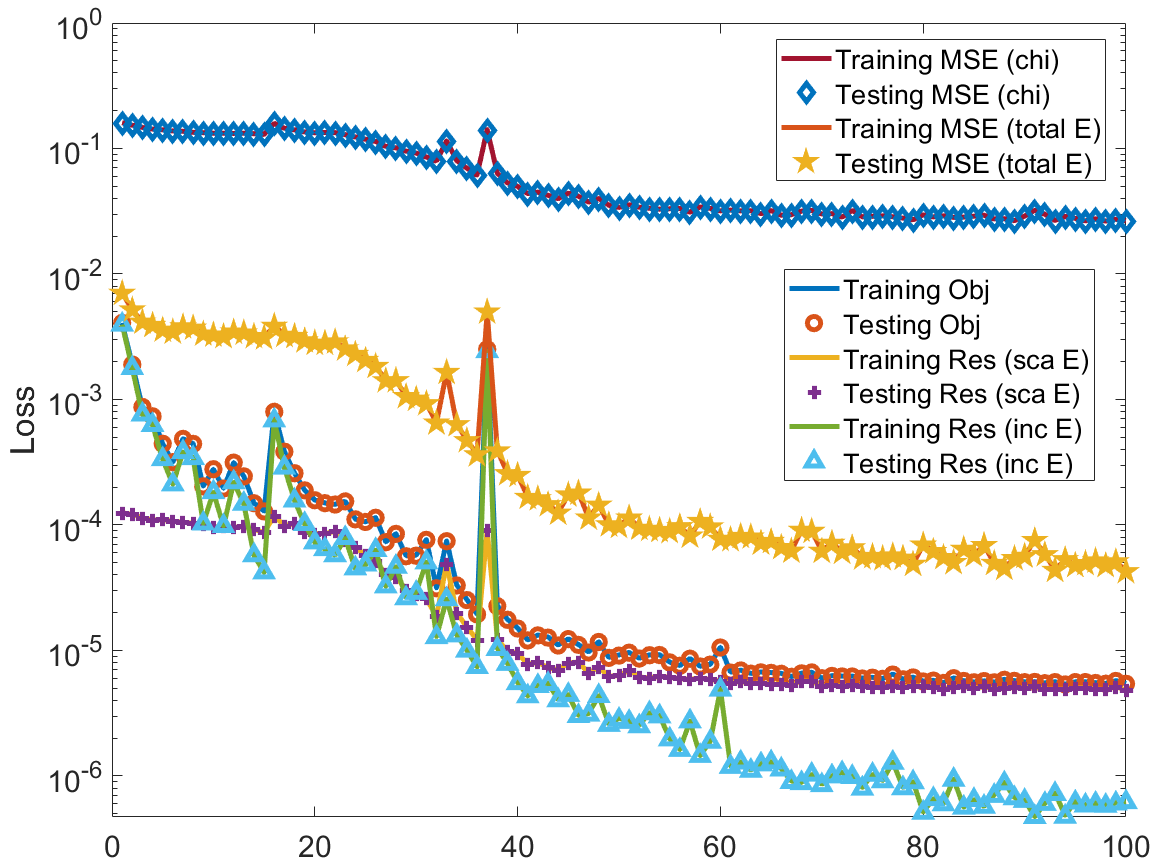}
	\caption{Convergence curve of unsupervised NeuralBIM. Training and testing Obj denote the curves of $obj_{su}$. Training and testing Res (sca E, inc E) denote the curves of $res_{\mathbf{E}^{s}}$, and $res_{\mathbf{E}^{i}}$. Training and testing MSE (chi, total E) denote the curves of $eval_{\boldsymbol{\chi}}$, and $eval_{\mathbf{E}^{t}}$.}
	\label{unsuloss}
\end{figure}
The convergence curves of unsupervised NeuralBIM are summarized in \fig{unsuloss}.
It is shown that $obj_{unsu}$, $res_{\mathbf{E}^{i}}$ and  $res_{\mathbf{E}^{s}}$ fall steadily as the training progresses.
They fluctuate several times due to the unsupervised learning scheme.
Since $res_{\mathbf{E}^{i}}$ and $res_{\mathbf{E}^{t}}$ cannot reflect the performance intuitively, the MSEs between the ground truth and the reconstructions are employed as the evaluation functions:
\begin{equation}
	\begin{aligned}
		eval_{\mathbf{E}^{t}} = &	\frac{1}{N_{{\mathbf{E}^{t}}^{\prime}}}||{\mathbf{E}^{t}}^{*} - {\mathbf{E}^{t}}^{\prime}||_{F}^2 \\
		eval_{\boldsymbol{\chi}} = & \frac{1}{N_{\boldsymbol{\chi}^{\prime}}}||\boldsymbol{\chi}^{*} - \boldsymbol{\chi}^{\prime}||_{F}^2  \,.
		\label{eq16}
	\end{aligned}
\end{equation}
The convergence curves of $eval_{\mathbf{E}^{t}}$ and $eval_{\boldsymbol{\chi}}$ are also plotted in \fig{unsuloss}.
The curves of $eval_{\mathbf{E}^{t}}$ and $eval_{\boldsymbol{\chi}}$ have the same trends as the ones of $obj_{unsu}$, which further validates the efficacy of the objective function based on the governing equations (\eq{eq3} and \eq{eq4}).
\begin{figure}
	\centering
	\subfigure[ ]{
		\centering
		\includegraphics[width=1\linewidth]{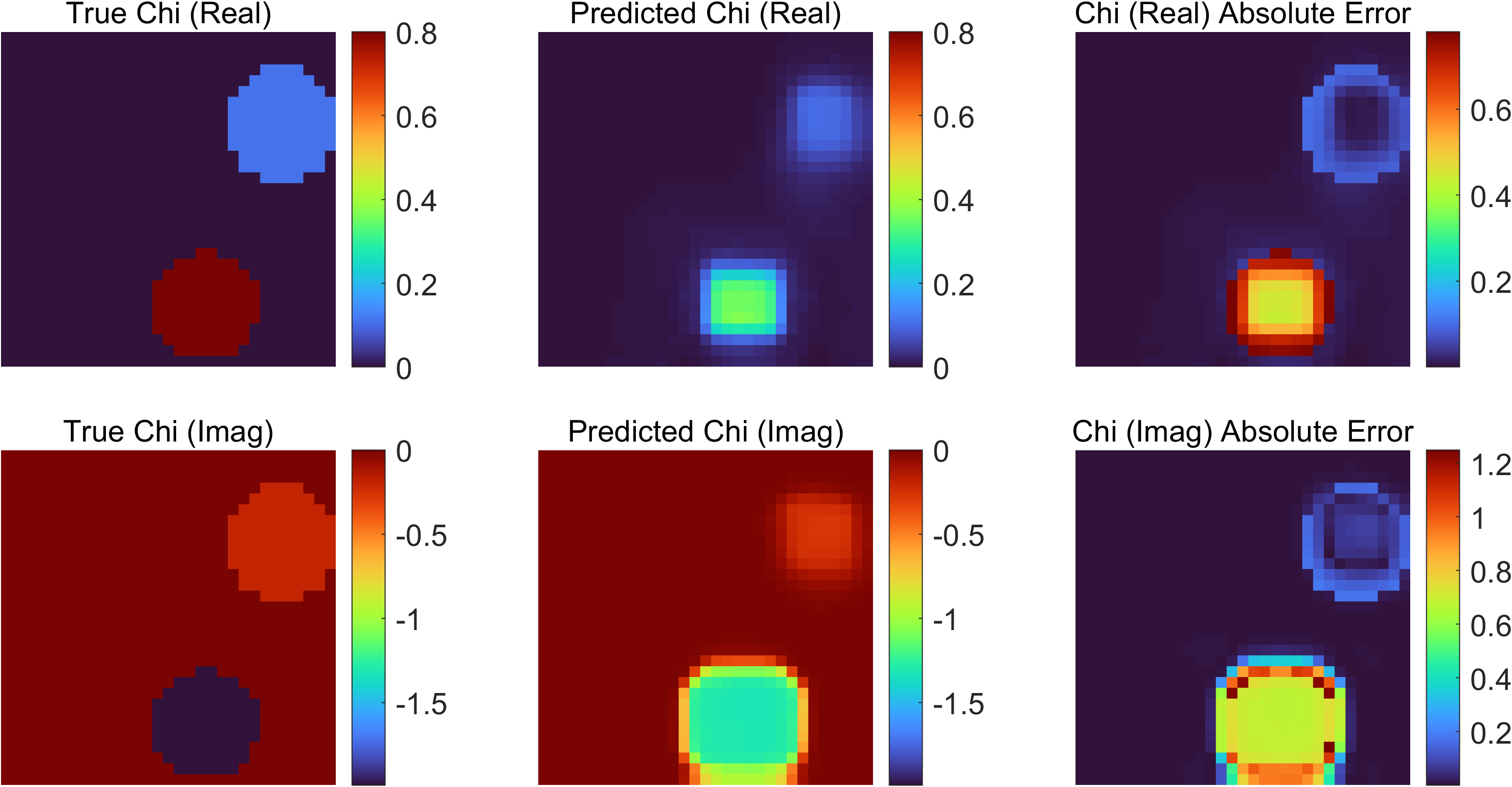}
	}
	\subfigure[ ]{
		\centering
		\includegraphics[width=1\linewidth]{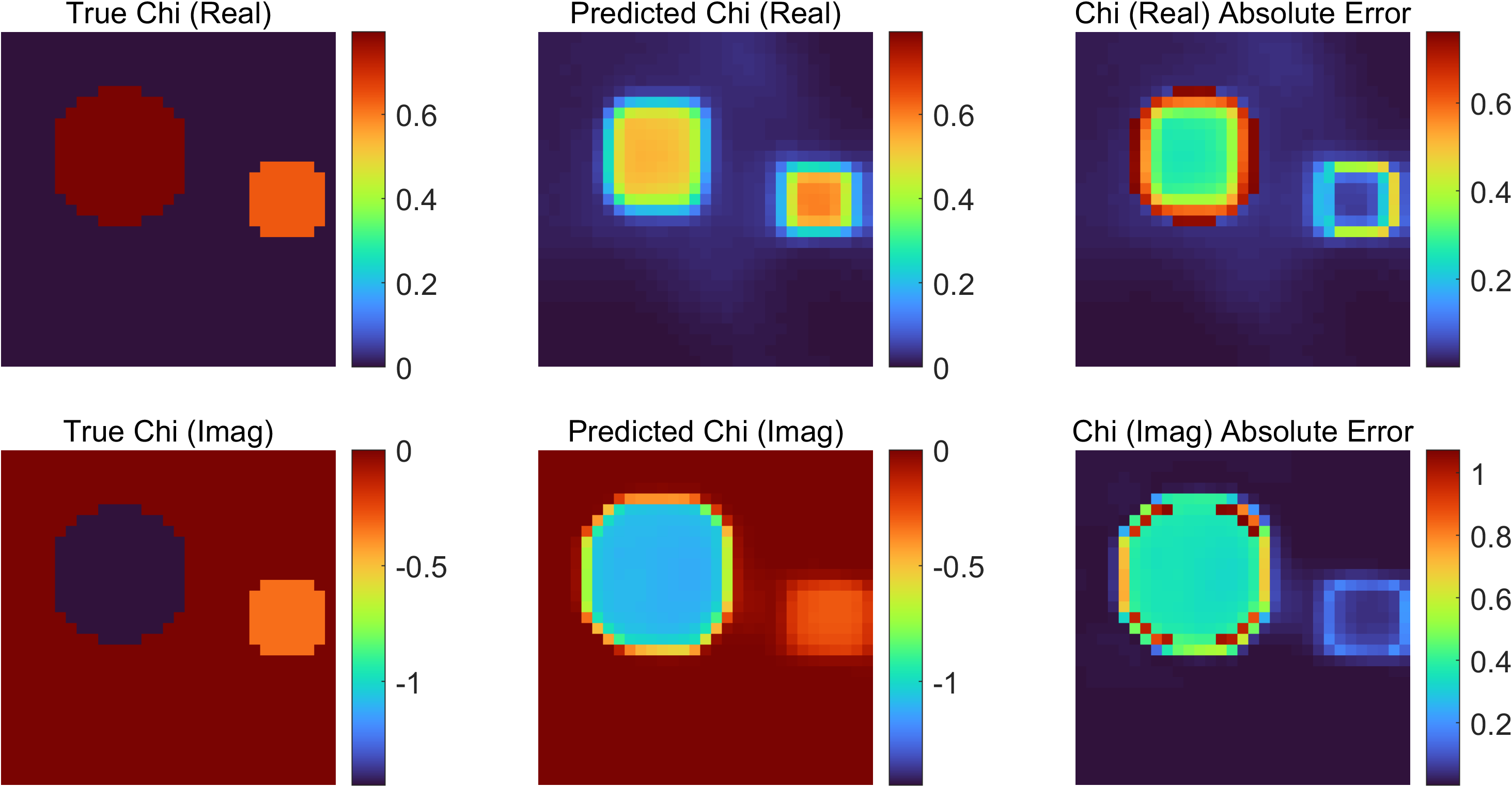}
	}
	\caption{Comparison between ground truth and scatterers inverted by unsupervised NeuralBIM. From left to right are: ground truth, reconstruction and their absolute error distribution.}
	\label{unsuchi}
\end{figure}
\begin{figure}
	\centering
	\subfigure[ ]{
		\centering
		\includegraphics[width=1\linewidth]{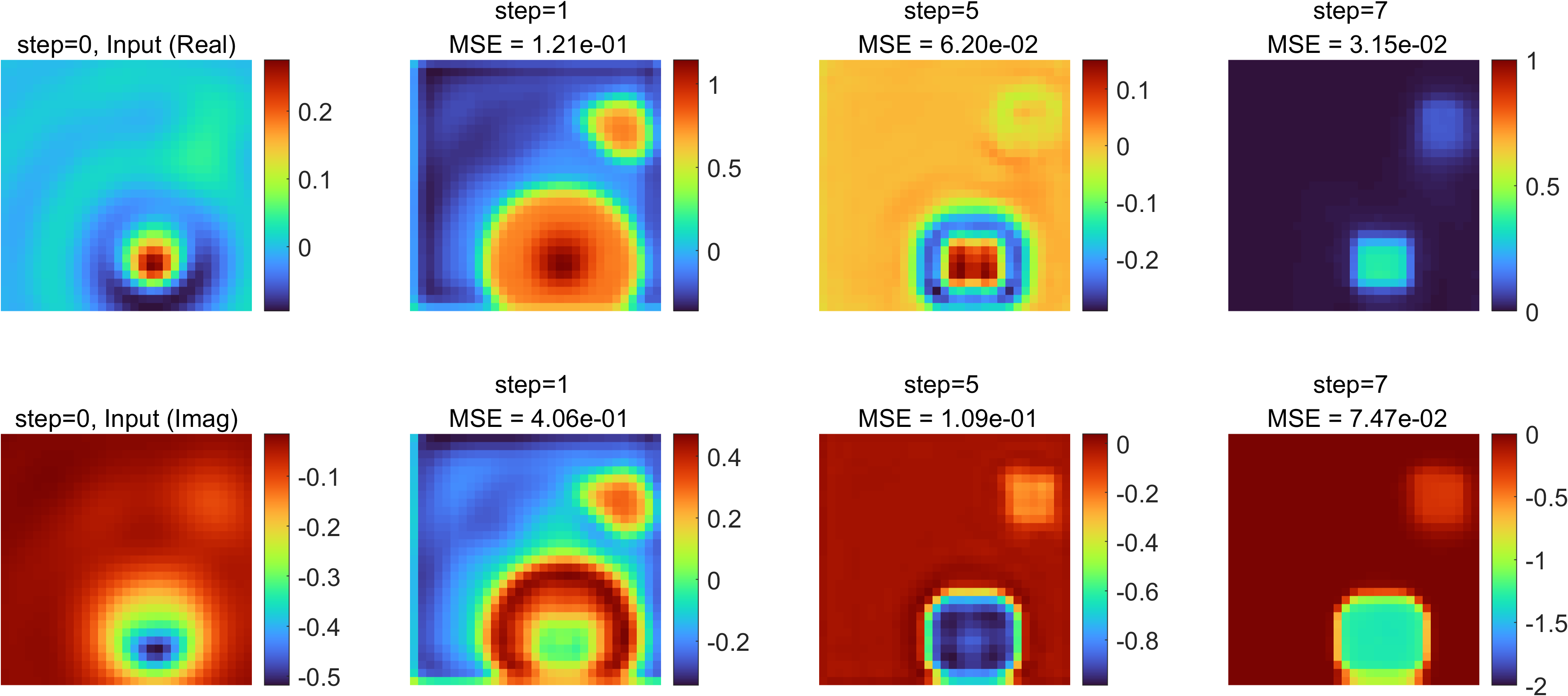}
	}
	\subfigure[ ]{
		\centering
		\includegraphics[width=1\linewidth]{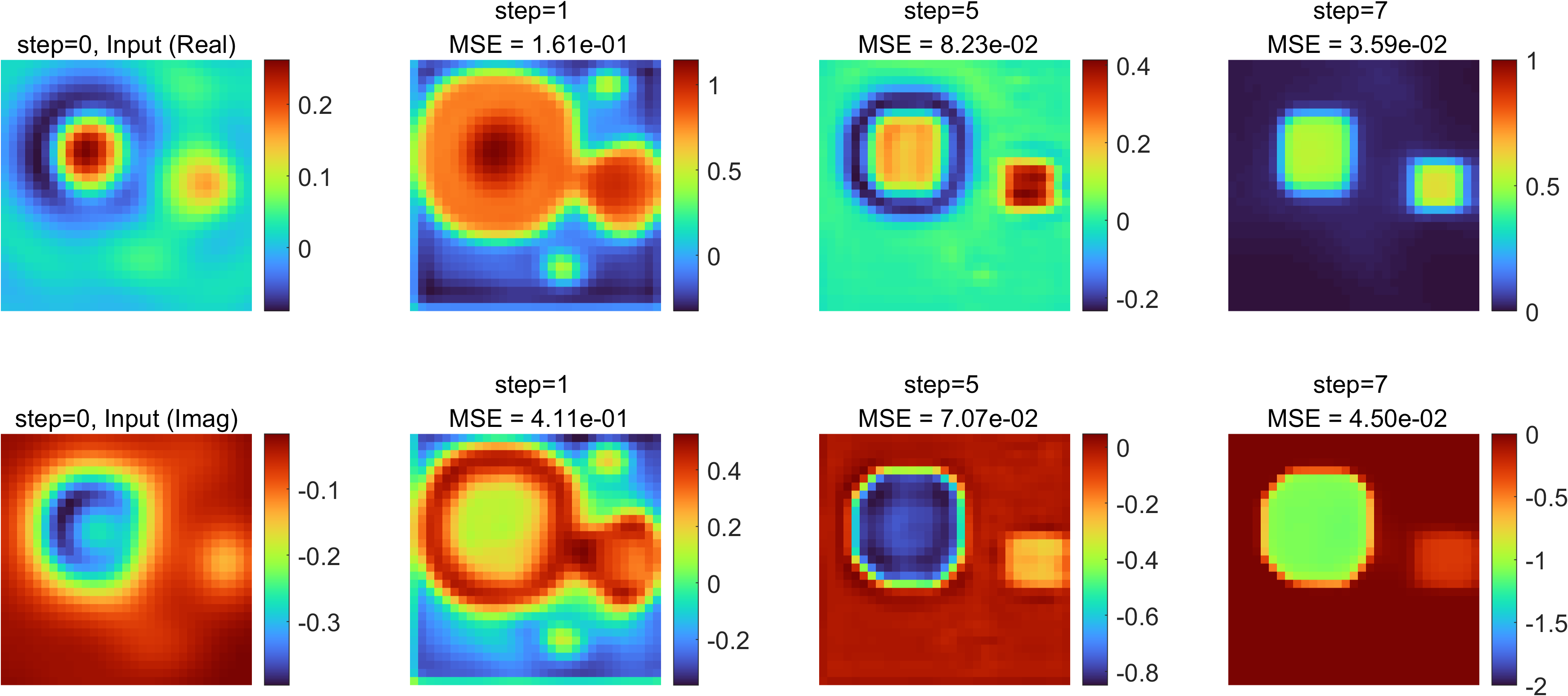}
	}
	\caption{Updated reconstructions of contrasts in the iterative computing of unsupervised NeuralBIM. From left to right are: initial guess, reconstructions at the first, fifth and seventh iteration. In each sub-panel, the first and second row show the real and imaginary parts.}
	\label{unsutotall}
\end{figure}
\begin{figure}
	
	\subfigure[ ]{
		\centering
	    \includegraphics[width=0.95\linewidth]{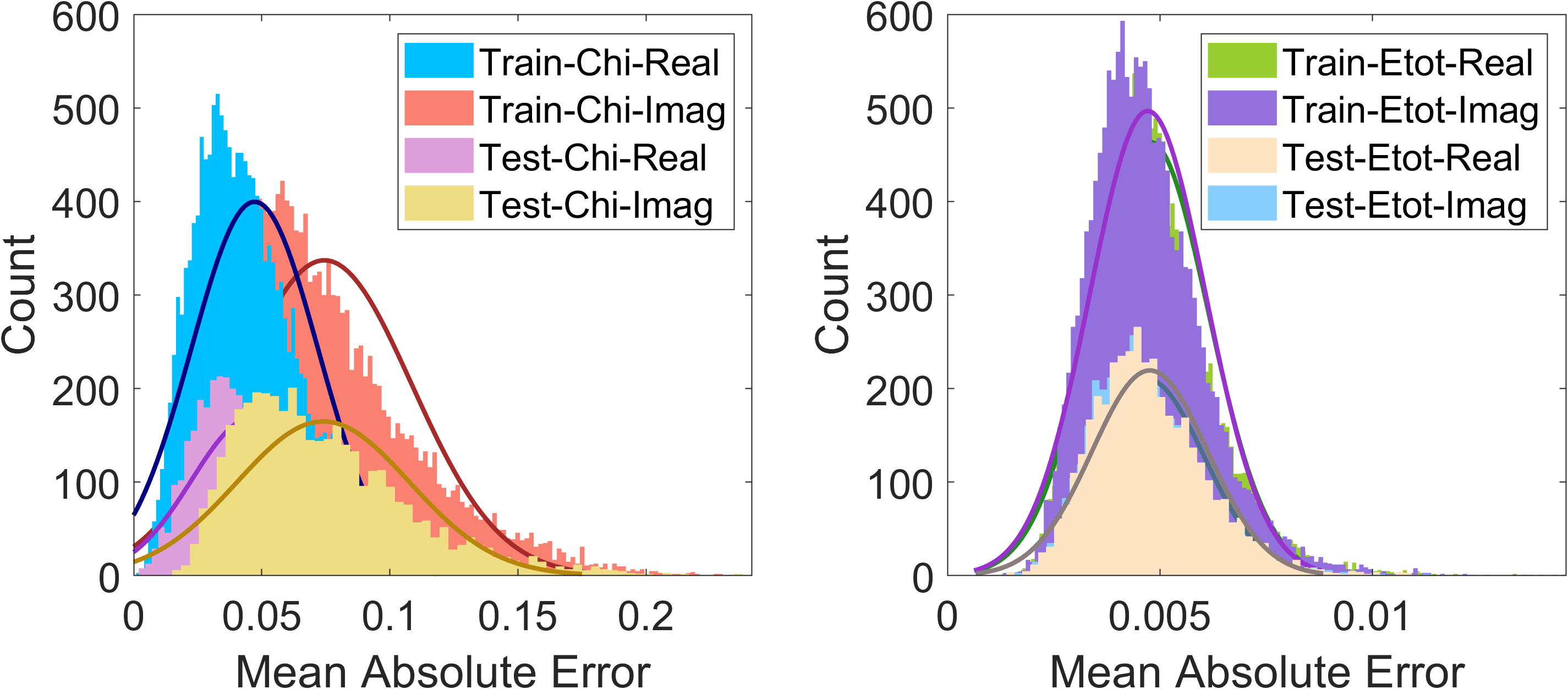}
        \label{unsuMAEall}}
	\subfigure[ ]{
		\centering
		\includegraphics[width=0.95\linewidth]{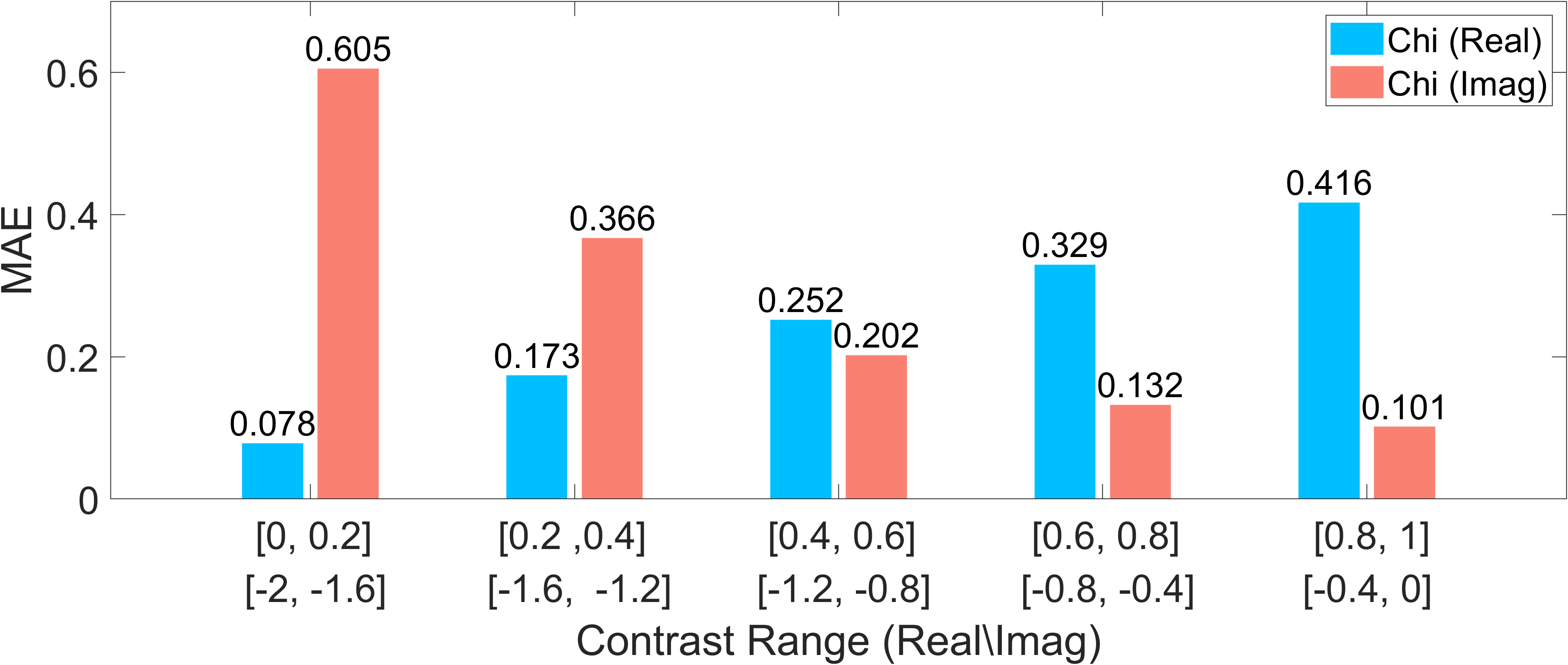}
	    \label{unsuMAEchi}}
	\caption{MAE histograms of contrasts and total fields inverted by unsupervised NeuralBIM. (a) is a histogram of the reconstruction MAE of each data sample in training and testing data sets. (b) is a histogram of the reconstruction MAE of each cylinder at different contrast ranges for all data samples. Chi, Etot, Real, Imag denote the contrast, total field, real part and imaginary part in the legend.}
	
\end{figure}
\par
\fig{unsuchi} demonstrates two reconstructed contrasts that are randomly chosen in the testing data set.
The inverted contrasts are in good agreement with the ground truth, and the absolute error distributions are in a low error level.
The TV regularization enables the homogenous background and the boundaries of scatterers in the reconstructions.

\fig{unsutotall} demonstrates the updated reconstructions of contrasts during the alternate update process.
The initial guesses of contrasts are generated by BP method.
The candidate contrasts are continuously modified with the increase of iterations.

\fig{unsuMAEall} charts the MAE histograms of inverted contrasts and total fields in the training and testing data sets.
The specific means and stds of MAE histograms are summarized in \tab{unsutab}.
The real parts of inverted scatterers have better precisions than the imaginary parts.
The MAE histograms of training and testing data have close means and stds, which is consistent with the convergence curve shown in \fig{unsuloss}.
\fig{unsuMAEchi} plots the reconstruction MAE of each cylinder at different contrast ranges for all data samples. The contrast range is divided in the same way as \fig{suMAEchi}. 
It can be observed that the MAEs grow with the increase of absolute values of contrasts.
In order to improve the performance on the cases of high contrasts, we can increase the proportion of high contrast samples in the training data set.

\begin{table}
	\renewcommand{\arraystretch}{1.3}
	\caption{MAE means and standard deviations of unsupervised NeuralBIM}
	\label{unsutab}
	\centering
	\begin{threeparttable}
		\begin{tabular}{ccc}
			\toprule
			Item \,\, \, & MAE-R\tnote{*} \, (mean/std) \,\,  & MAE-I\tnote{+} \, (mean/std) \,\,\,\,\, \\
			\midrule
			Contrast (train)\tnote{1} & $0.0472/0.0248$ & $0.0745/0.0341$\\
			Contrast (test)\tnote{2} & $0.0465/0.0239$ & $0.0740/0.0337$ \\
			Total field (train)\tnote{3} & $0.0047/0.0014$ & $0.0048/0.0014$ \\
			Total field (test)\tnote{4} & $0.00047/0.0013$ & $0.0048/0.0014$ \\
			\bottomrule
		\end{tabular}
		\begin{tablenotes}
			\setlength{\multicolsep}{0cm}
			\begin{multicols}{2}
				\item[1] contrasts in  training data set
				\item[2] contrasts in  testing data set
				\item[3] total fields in  training data set
				\item[4] total fields in  testing data set \\
				\item[*] MAE of real part 	
				\item[+] MAE of imaginary part
			\end{multicols}
		\end{tablenotes}
	\end{threeparttable}
\end{table}
\begin{figure}
	\centering
	\includegraphics[width=1\linewidth]{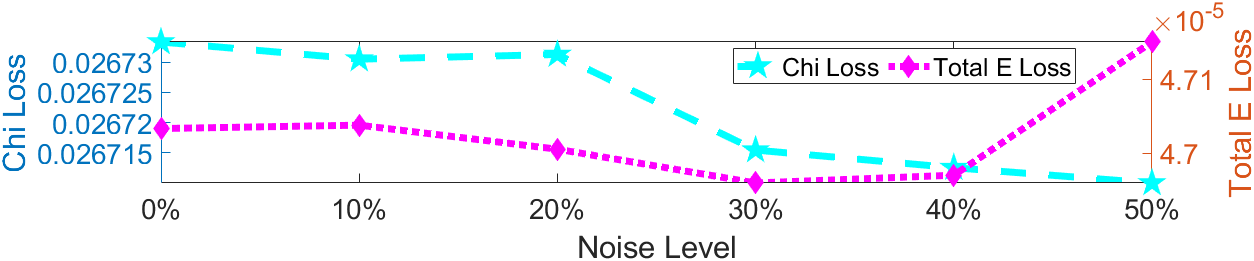}
	\caption{Losses of contrasts and total fields reconstructed by unsupervised NeuralBIM under different noise levels. Chi and total E denote contrast and total field.}
	\label{noiseloss-unsu}
\end{figure}
\begin{figure}
	\centering
	\includegraphics[width=1\linewidth]{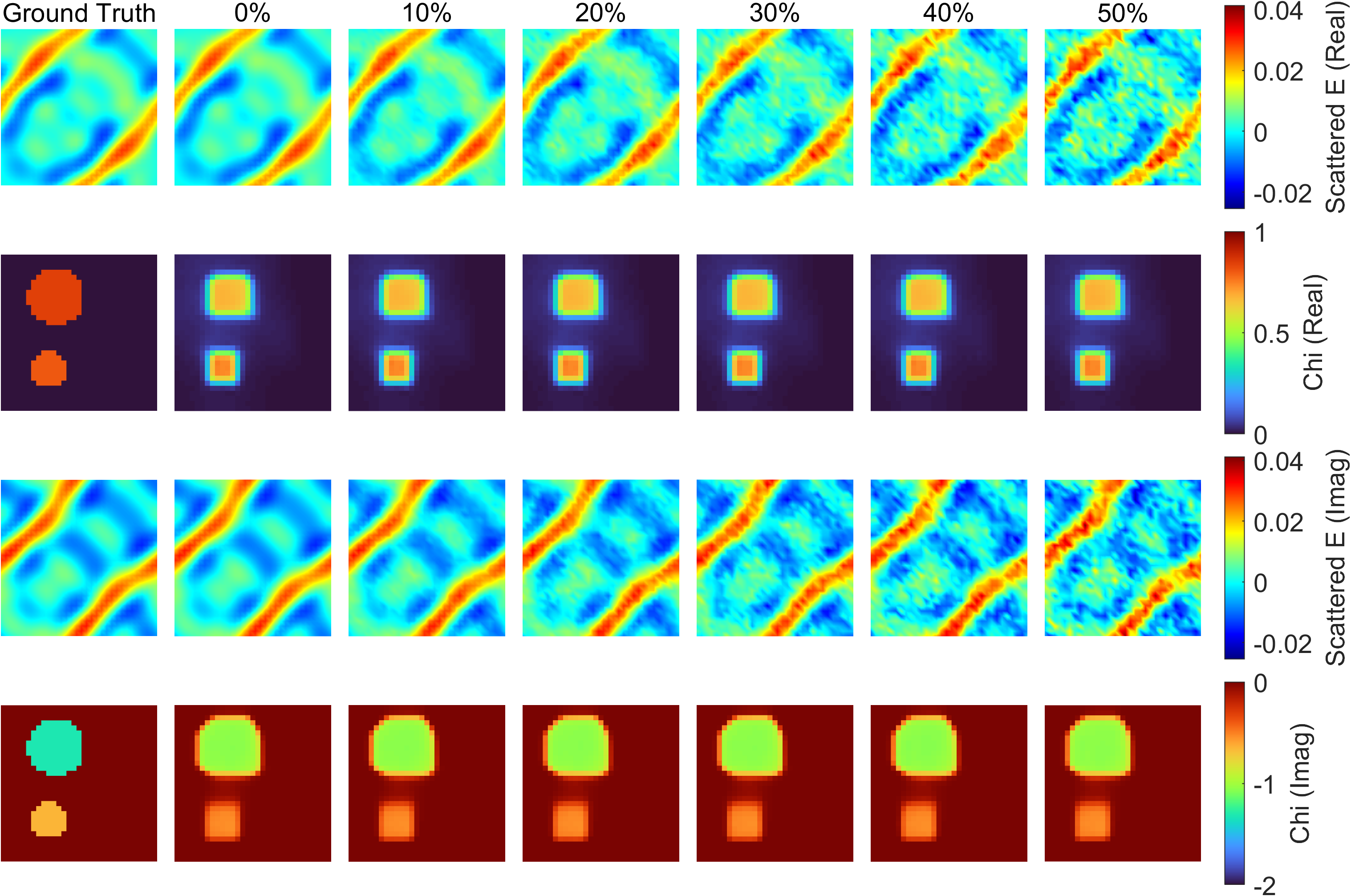}
	\caption{Reconstructed scatterers of unsupervised NeuralBIM under different noise levels. From left to right are ground truth, reconstructions at the noise level of 0\%, 10\%, 20\%, 30\%, 40\%, 50\%.}
	\label{noisetest-unsu}
\end{figure}
\begin{figure}
	\centering
	\includegraphics[width=1\linewidth]{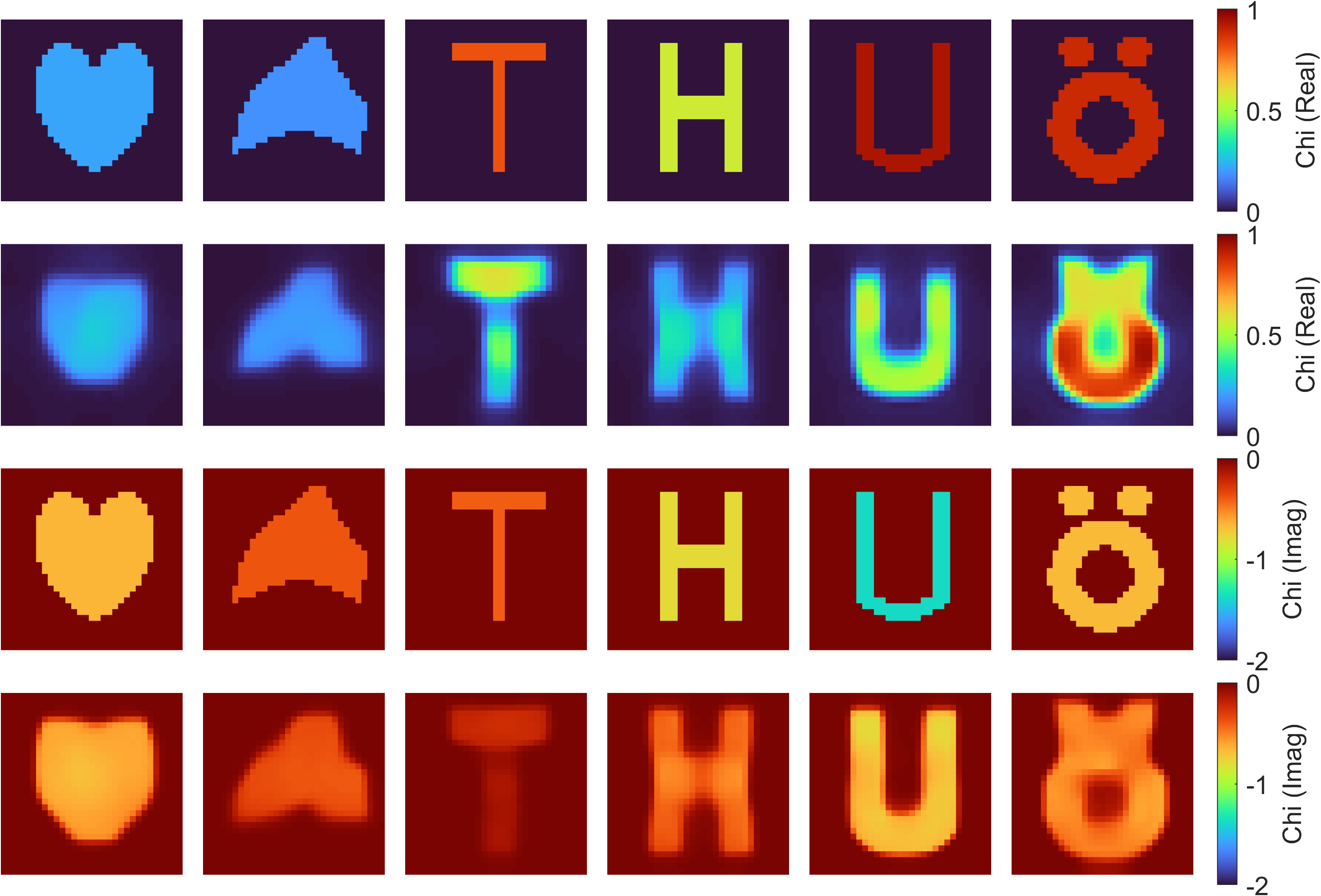}
	\caption{Verification of generalization ability of unsupervised NeuralBIM on different scatterer shapes.}
	\label{geometrytest-unsu}
\end{figure}
\par 
The performance of unsupervised NeuralBIM is then evaluated under different noise levels of scattered fields.
A total of 20 data samples are selected from the testing data set and the Gaussian white noise is added to the field data for validation.
The curves of $eval_{\mathbf{E}^{t}}$ and $eval_{\boldsymbol{\chi}}$ are plotted in \fig{noiseloss-unsu}.
It can be observed that both $eval_{\mathbf{E}^{t}}$ and $eval_{\boldsymbol{\chi}}$ are stable under different noise levels.
The anti-noise capability of unsupervised NeuralBIM is verified.
The inverted scatterers at the noise level from 0\% to 50\% are compared in \fig{noisetest-unsu}.
The inverted scatterers demonstrates a good and stable image quality.

The generalization ability on scatterer shapes is then verified.
The shapes of scatterers are totally different from the training ones.
The corresponding reconstructions are demonstrated in \fig{geometrytest-unsu}.
The inverted scatterers are in good agreement with the ground truth with clear boundaries.
\subsection{Comparison With TBIM}
\begin{figure}
	\centering
	\includegraphics[width=1\linewidth]{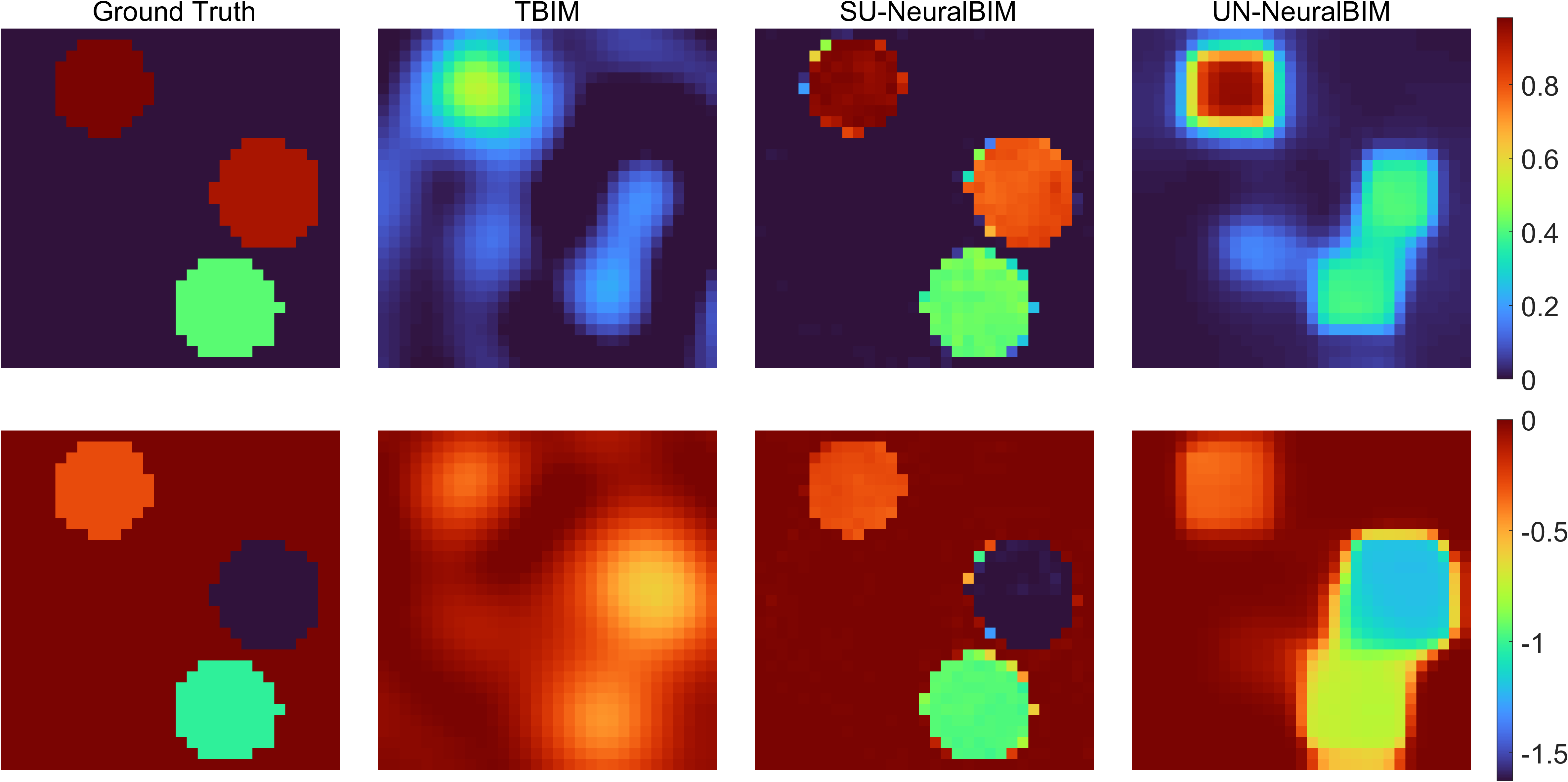}
	\caption{Comparison between reconstructions of TBIM, supervised NeuralBIM and unsupervised NeuralBIM. From left to right are ground truth, reconstructions of TBIM, supervised NeuralBIM and unsupervised NeuralBIM. The first row shows the real parts and the second shows the imaginary parts.}
	\label{compareBIM}
\end{figure}

The performances of TBIM, supervised and unsupervised NeuralBIM are compared in this section.
NeuralBIM is assumed to stop at the $7$-th iteration and TBIM converges at the $14$-th iteration.
The TBIM is computed on Intel(R) Core(TM) i5-9600K CPU @ 3.70GHz.
Besides, the initial guess of TBIM is also generated by BP method.
The testing example assumes that three cylinders are located in $D$.
The real and imaginary parts of their contrasts are in $[0, 1]$ and $[-2, 0]$ respectively.
Note that the testing example is unseen during the training of both supervised and unsupervised NeuralBIM.
\fig{compareBIM} illustrates the comparison between reconstructions of TBIM, supervised and unsupervised NeuralBIM.
TBIM can locate the positions of three cylinders, but it cannot capture the accurate shapes of scatterers with some artifacts.
Supervised NeuralBIM demonstrates the best performance.
The inverted result is high-resolution and the difference from the ground truth primarily lies on the boundaries of cylinders.
Unsupervised NeuralBIM can also locate the cylinders and provide better shapes of scatterers than TBIM.
The detailed comparison of performance is summarized in \tab{CompTab}.
Although supervised and unsupervised NeuralBIM needs approximately 41 hours and 45 hours for offline training respectively, they demonstrates a significant reduction in online computing time compared to TBIM. 
The MAEs between ground truth and reconstructions of these methods further validate the aforementioned observations derived from \fig{compareBIM}.
\begin{table}
	\centering
	\caption{Performance of TBIM, supervised and unsupervised Neural BIM}
	\label{CompTab}
	\begin{threeparttable}
	\begin{tabular}{cccc}
		\hline
		  & TBIM & SU-NeuralBIM\tnote{*} &  UN-NeuralBIM\tnote{+}\\ \hline
		Training time & 0 & $\approx 41h$ & $\approx 45h$ \\
		Inference time & $4.25s$ & $0.046s$ & $0.046s$ \\ 
		MAE (Real/Imag)& $0.177/0.216$ & $0.0297/0.0280$ & $0.113/0.132$ \\ 
		\hline
	\end{tabular}
	\begin{tablenotes}
	\setlength{\multicolsep}{0cm}
	\begin{multicols}{2}
		\item[*] Supervised NeuralBIM 	
		\item[+] Unsupervised NeuralBIM
	\end{multicols}
    \end{tablenotes}
	\end{threeparttable}
\end{table}
\subsection{Experimental Data Inversion}
\label{expsection}
\begin{table}
	\centering
	\caption{Training Settings in Experimental Data Inversion}
	\label{Tab1}
	\begin{tabular}{cccc}
		\hline
		Cylinder & Radius & Contrast (Real) & Contrast (Imag)\\ \hline
		A & 0.03-0.045$m$ & 0.2-0.8 & 0 \\
		B & 0.01-0.02$m$ & 1.5-2.3 & 0 \\ \hline
	\end{tabular}
\end{table}
\begin{figure}
	\centering
	\subfigure[]
	{\includegraphics[width=0.9\linewidth]{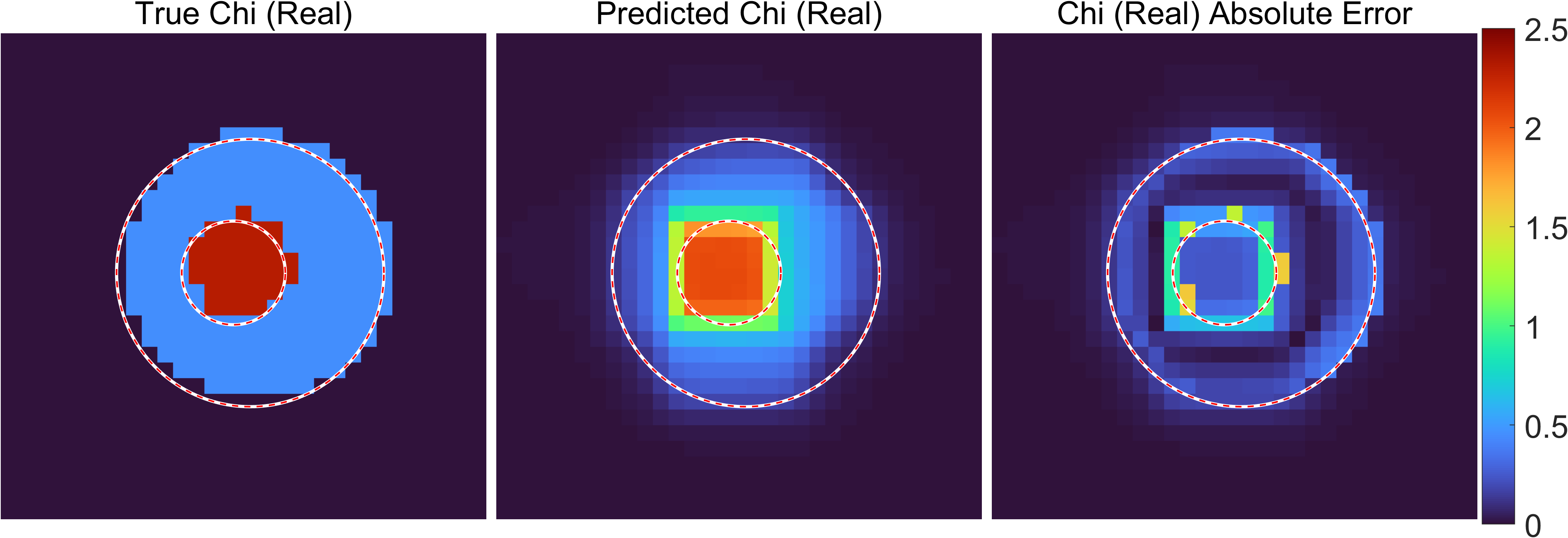}}
	\subfigure[]
	{\includegraphics[width=1\linewidth]{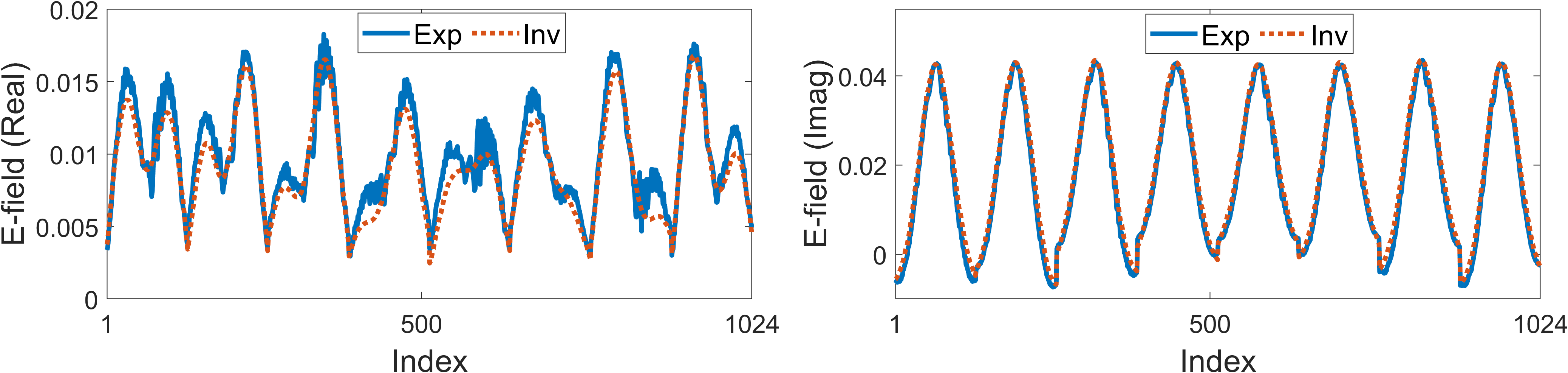}}	
	\subfigure[]
	{\includegraphics[width=0.9\linewidth]{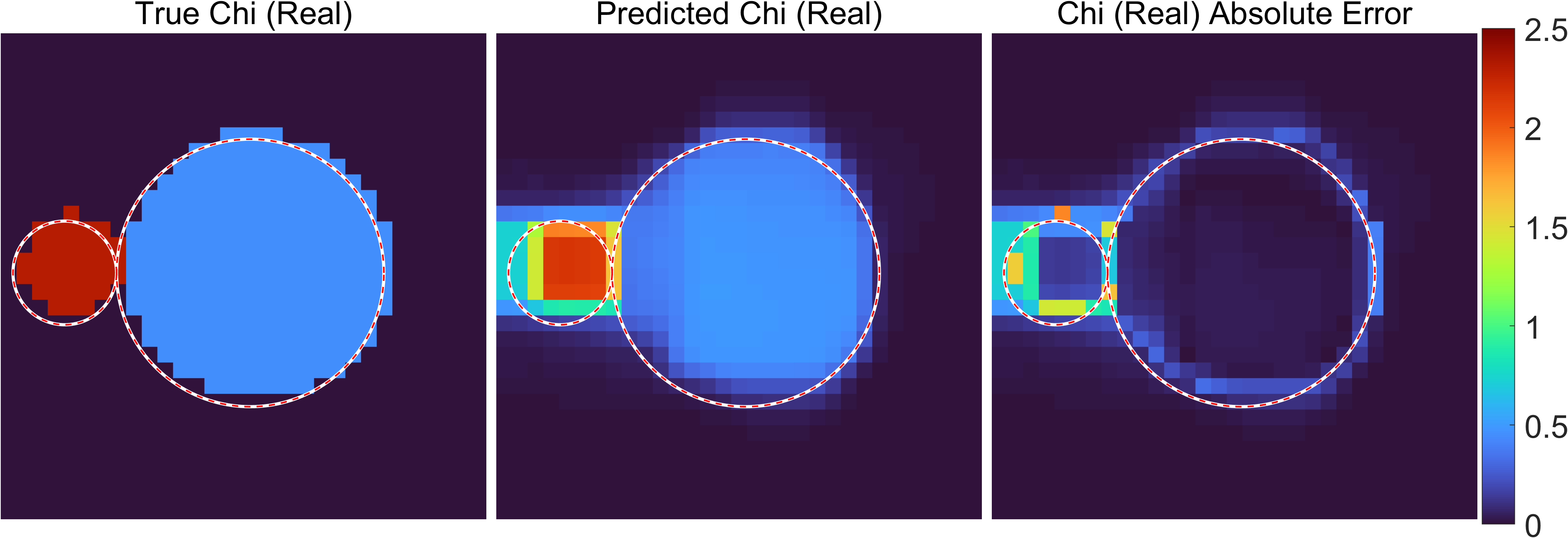}}
	\subfigure[]
	{\includegraphics[width=1\linewidth]{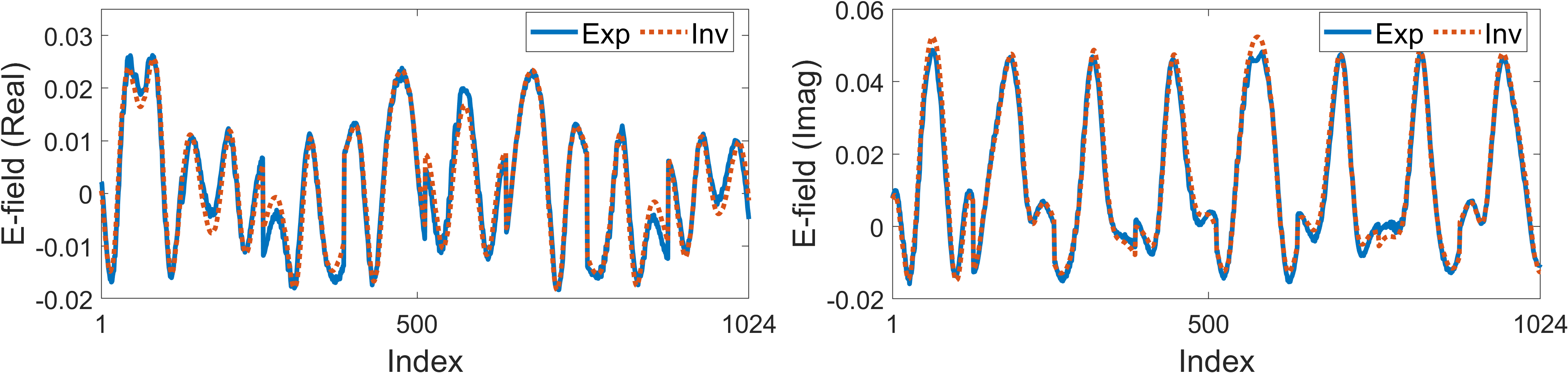}}
	\caption{Experimental data inversion of \textit{FoamDielInt} model [(a), (b)] and \textit{FoamDielExt} model [(c), (d)]. (a) and (c): the ground truth, reconstruction, and their corresponding absolute error distribution. (b) and (d): the inverted (red dashed line) and measured (blue solid line) scattered field.}
	\label{FresnelModel}
\end{figure}
In this section, we validate unsupervised NeuralBIM with experimental data inversion.
The experimental data is published by Institut Fresnel\cite{geffrin2005free}.
The TM polarized measured data of \textit{FoamDielInt} and \textit{FoamDielExt} models are selected for inversion.
The frequency of measured data is 3 GHz. 
The measurement is performed with 8 transmitters and 241 receivers in \cite{geffrin2005free}.
The measurement configuration is different from the synthetic one.
The Green's functions in \eq{eq3} and \eq{eq4} are changed as a result.
Because the Green's functions are incorporated, NeuralBIM needs to be re-trained when the measurement configuration changes.

The training configuration has 8 transmitters and downsamples 241 receivers to 128 receivers.
The training data set of 16000 samples is generated by assuming two cylinders are randomly located inside the DoI.
\tab{Tab1} summarizes the geometries and material properties of cylinders.
The measured data takes the size of $128 \times 8$ and they are reshaped into $32 \times32$ for training.

After the training is finished, unsupervised NeuralBIM is employed to invert the measurement data of \textit{FoamDielInt} and \textit{FoamDielExt} models. 
\fig{FresnelModel} shows the reconstructions and the corresponding inverted scattered fields.
The reconstructed geometries and material properties of scatterers agree well with the ground truth.
Then the scattered fields are calculated based on the reconstructed results.
The inverted scattered fields also show a good agreement with the measured ones, which further validate unsupervised NeuralBIM.

\section{Conclusions}
In this paper, neural Born iterative method is proposed to solve 2D inverse scattering problems.
Inspired by PhiSRL, NeuralBIM emulates the computational process of TBIM by applying CNNs to learn the parametric functions of alternate update rules.
Both supervised and unsupervised learning schemes of NeuralBIM are presented in this paper, and they all demonstrate better performance than TBIM.
Numerical and experimental results validate the efficacy of the proposed NeuralBIM.
The anti-noise capability and generalization ability of NeuralBIM are also verified.

This paper investigate a potential way to incorporate deep learning techniques into traditional computational electromagnetic algorithms.
Additionally, the feasibility is also verified that the physics-incorporated deep neural network can be trained via an unsupervised process to perform reliable computations by applying governing equations as its objective function.

% references section
\bibliographystyle{IEEEtran}
\bibliography{IEEEabrv,ref}

% Generated by IEEEtran.bst, version: 1.14 (2015/08/26)
\begin{thebibliography}{10}
\providecommand{\url}[1]{#1}
\csname url@samestyle\endcsname
\providecommand{\newblock}{\relax}
\providecommand{\bibinfo}[2]{#2}
\providecommand{\BIBentrySTDinterwordspacing}{\spaceskip=0pt\relax}
\providecommand{\BIBentryALTinterwordstretchfactor}{4}
\providecommand{\BIBentryALTinterwordspacing}{\spaceskip=\fontdimen2\font plus
\BIBentryALTinterwordstretchfactor\fontdimen3\font minus
  \fontdimen4\font\relax}
\providecommand{\BIBforeignlanguage}[2]{{%
\expandafter\ifx\csname l@#1\endcsname\relax
\typeout{** WARNING: IEEEtran.bst: No hyphenation pattern has been}%
\typeout{** loaded for the language `#1'. Using the pattern for}%
\typeout{** the default language instead.}%
\else
\language=\csname l@#1\endcsname
\fi
#2}}
\providecommand{\BIBdecl}{\relax}
\BIBdecl

\bibitem{chen2018computational}
X.~Chen, \emph{{Computational methods for electromagnetic inverse
  scattering}}.\hskip 1em plus 0.5em minus 0.4em\relax John Wiley \& Sons,
  2018.

\bibitem{salucci2016real}
M.~Salucci, N.~Anselmi, G.~Oliveri, P.~Calmon, R.~Miorelli, C.~Reboud, and
  A.~Massa, ``{Real-time NDT-NDE through an innovative adaptive partial least
  squares SVR inversion approach},'' \emph{IEEE Transactions on Geoscience and
  Remote Sensing}, vol.~54, no.~11, pp. 6818--6832, 2016.

\bibitem{abubakar2002imaging}
A.~Abubakar, P.~M. Van~den Berg, and J.~J. Mallorqui, ``{Imaging of biomedical
  data using a multiplicative regularized contrast source inversion method},''
  \emph{IEEE Transactions on Microwave Theory and Techniques}, vol.~50, no.~7,
  pp. 1761--1771, 2002.

\bibitem{pastorino2018microwave}
M.~Pastorino and A.~Randazzo, \emph{{Microwave Imaging Methods and
  Applications}}.\hskip 1em plus 0.5em minus 0.4em\relax Artech House, 2018.

\bibitem{abubakar20082}
A.~Abubakar, T.~Habashy, V.~Druskin, L.~Knizhnerman, and D.~Alumbaugh, ``{2.5 D
  forward and inverse modeling for interpreting low-frequency electromagnetic
  measurements},'' \emph{Geophysics}, vol.~73, no.~4, pp. F165--F177, 2008.

\bibitem{salucci2016multifrequency}
M.~Salucci, L.~Poli, N.~Anselmi, and A.~Massa, ``{Multifrequency particle swarm
  optimization for enhanced multiresolution GPR microwave imaging},''
  \emph{IEEE Transactions on Geoscience and Remote Sensing}, vol.~55, no.~3,
  pp. 1305--1317, 2016.

\bibitem{slaney1984limitations}
M.~Slaney, A.~C. Kak, and L.~E. Larsen, ``{Limitations of imaging with
  first-order diffraction tomography},'' \emph{IEEE transactions on microwave
  theory and techniques}, vol.~32, no.~8, pp. 860--874, 1984.

\bibitem{devaney1981inverse}
A.~Devaney, ``{Inverse-scattering theory within the Rytov approximation},''
  \emph{Optics letters}, vol.~6, no.~8, pp. 374--376, 1981.

\bibitem{belkebir2005superresolution}
K.~Belkebir, P.~C. Chaumet, and A.~Sentenac, ``{Superresolution in total
  internal reflection tomography},'' \emph{JOSA A}, vol.~22, no.~9, pp.
  1889--1897, 2005.

\bibitem{wang1989iterative}
Y.~Wang and W.~C. Chew, ``{An iterative solution of the two-dimensional
  electromagnetic inverse scattering problem},'' \emph{International Journal of
  Imaging Systems and Technology}, vol.~1, no.~1, pp. 100--108, 1989.

\bibitem{chew1990reconstruction}
W.~C. Chew and Y.-M. Wang, ``{Reconstruction of two-dimensional permittivity
  distribution using the distorted Born iterative method},'' \emph{IEEE
  transactions on medical imaging}, vol.~9, no.~2, pp. 218--225, 1990.

\bibitem{van2001contrast}
P.~Van~den Berg and A.~Abubakar, ``{Contrast source inversion method: State of
  art},'' \emph{Progress in Electromagnetics Research}, vol.~34, pp. 189--218,
  2001.

\bibitem{zakaria2010finite}
A.~Zakaria, C.~Gilmore, and J.~LoVetri, ``{Finite-element contrast source
  inversion method for microwave imaging},'' \emph{Inverse Problems}, vol.~26,
  no.~11, p. 115010, 2010.

\bibitem{chen2009subspace}
X.~Chen, ``{Subspace-based optimization method for solving inverse-scattering
  problems},'' \emph{IEEE Transactions on Geoscience and Remote Sensing},
  vol.~48, no.~1, pp. 42--49, 2009.

\bibitem{dorn2006level}
O.~Dorn and D.~Lesselier, ``{Level set methods for inverse scattering},''
  \emph{Inverse Problems}, vol.~22, no.~4, p. R67, 2006.

\bibitem{mojabi2009overview}
P.~Mojabi and J.~LoVetri, ``{Overview and classification of some regularization
  techniques for the Gauss-Newton inversion method applied to inverse
  scattering problems},'' \emph{IEEE Transactions on Antennas and Propagation},
  vol.~57, no.~9, pp. 2658--2665, 2009.

\bibitem{golub1999tikhonov}
G.~H. Golub, P.~C. Hansen, and D.~P. O'Leary, ``{Tikhonov regularization and
  total least squares},'' \emph{SIAM journal on matrix analysis and
  applications}, vol.~21, no.~1, pp. 185--194, 1999.

\bibitem{van1995total}
P.~M. van~den Berg and R.~E. Kleinman, ``{A total variation enhanced modified
  gradient algorithm for profile reconstruction},'' \emph{Inverse Problems},
  vol.~11, no.~3, p.~L5, 1995.

\bibitem{candes2008introduction}
E.~J. Cand{\`e}s and M.~B. Wakin, ``{An introduction to compressive
  sampling},'' \emph{IEEE signal processing magazine}, vol.~25, no.~2, pp.
  21--30, 2008.

\bibitem{oliveri2017compressive}
G.~Oliveri, M.~Salucci, N.~Anselmi, and A.~Massa, ``{Compressive Sensing as
  Applied to Inverse Problems for Imaging: Theory, applications, current
  trends, and open challenges.}'' \emph{IEEE Antennas and Propagation
  Magazine}, vol.~59, no.~5, pp. 34--46, 2017.

\bibitem{guo2015microwave}
L.~Guo and A.~Abbosh, ``{Microwave stepped frequency head imaging using
  compressive sensing with limited number of frequency steps},'' \emph{IEEE
  Antennas and Wireless Propagation Letters}, vol.~14, pp. 1133--1136, 2015.

\bibitem{oliveri2019compressive}
G.~Oliveri, L.~Poli, N.~Anselmi, M.~Salucci, and A.~Massa, ``{Compressive
  sensing-based Born iterative method for tomographic imaging},'' \emph{IEEE
  Transactions on Microwave Theory and Techniques}, vol.~67, no.~5, pp.
  1753--1765, 2019.

\bibitem{pan2012compressive}
L.~Pan, X.~Chen, and S.~P. Yeo, ``{A compressive-sensing-based phaseless
  imaging method for point-like dielectric objects},'' \emph{IEEE transactions
  on antennas and propagation}, vol.~60, no.~11, pp. 5472--5475, 2012.

\bibitem{guo2020application}
L.~Guo, M.~Li, S.~Xu, and F.~Yang, ``{Application of Stochastic Gradient
  Descent Technique for Method of Moments},'' in \emph{2020 IEEE International
  Conference on Computational Electromagnetics (ICCEM)}.\hskip 1em plus 0.5em
  minus 0.4em\relax IEEE, 2020, pp. 97--98.

\bibitem{shan2021application}
T.~Shan, R.~Guo, M.~Li, F.~Yang, S.~Xu, and L.~Liang, ``{Application of
  Multitask Learning for 2-D Modeling of Magnetotelluric Surveys: TE Case},''
  \emph{IEEE Transactions on Geoscience and Remote Sensing}, vol.~60, pp. 1--9,
  2021.

\bibitem{shan2020study}
T.~Shan, W.~Tang, X.~Dang, M.~Li, F.~Yang, S.~Xu, and J.~Wu, ``{Study on a fast
  solver for Poisson’s equation based on deep learning technique},''
  \emph{IEEE Transactions on Antennas and Propagation}, vol.~68, no.~9, pp.
  6725--6733, 2020.

\bibitem{massa2019dnns}
A.~Massa, D.~Marcantonio, X.~Chen, M.~Li, and M.~Salucci, ``{DNNs as applied to
  electromagnetics, antennas, and propagation—A review},'' \emph{IEEE
  Antennas and Wireless Propagation Letters}, vol.~18, no.~11, pp. 2225--2229,
  2019.

\bibitem{shan2020coding}
T.~Shan, X.~Pan, M.~Li, S.~Xu, and F.~Yang, ``{Coding programmable metasurfaces
  based on deep learning techniques},'' \emph{IEEE Journal on Emerging and
  Selected Topics in Circuits and Systems}, vol.~10, no.~1, pp. 114--125, 2020.

\bibitem{shan2021phase}
T.~Shan, M.~Li, S.~Xu, and F.~Yang, ``{Phase Synthesis of Beam-Scanning
  Reflectarray Antenna Based on Deep Learning Technique},'' \emph{Progress In
  Electromagnetics Research}, vol. 172, pp. 41--49, 2021.

\bibitem{chen2020review}
X.~Chen, Z.~Wei, M.~Li, and P.~Rocca, ``{A review of deep learning approaches
  for inverse scattering problems (invited review)},'' \emph{Progress In
  Electromagnetics Research}, vol. 167, pp. 67--81, 2020.

\bibitem{salucci2022artificial}
M.~Salucci, M.~Arrebola, T.~Shan, and M.~Li, ``{Artificial Intelligence: New
  Frontiers in Real--Time Inverse Scattering and Electromagnetic Imaging},''
  \emph{IEEE Transactions on Antennas and Propagation}, 2022.

\bibitem{guo2019supervised}
R.~Guo, X.~Song, M.~Li, F.~Yang, S.~Xu, and A.~Abubakar, ``{Supervised descent
  learning technique for 2-D microwave imaging},'' \emph{IEEE Transactions on
  Antennas and Propagation}, vol.~67, no.~5, pp. 3550--3554, 2019.

\bibitem{xu2020deep}
K.~Xu, L.~Wu, X.~Ye, and X.~Chen, ``{Deep learning-based inversion methods for
  solving inverse scattering problems with phaseless data},'' \emph{IEEE
  Transactions on Antennas and Propagation}, vol.~68, no.~11, pp. 7457--7470,
  2020.

\bibitem{zhang2022accelerating}
R.~Zhang, Q.~Sun, Y.~Mao, L.~Cui, Y.~Jia, W.-F. Huang, M.~Ahmadian, and Q.~H.
  Liu, ``{Accelerating Hydraulic Fracture Imaging by Deep Transfer Learning},''
  \emph{IEEE Transactions on Antennas and Propagation}, 2022.

\bibitem{song2021electromagnetic}
R.~Song, Y.~Huang, K.~Xu, X.~Ye, C.~Li, and X.~Chen, ``{Electromagnetic inverse
  scattering with perceptual generative adversarial networks},'' \emph{IEEE
  Transactions on Computational Imaging}, vol.~7, pp. 689--699, 2021.

\bibitem{li2018deepnis}
L.~Li, L.~G. Wang, F.~L. Teixeira, C.~Liu, A.~Nehorai, and T.~J. Cui,
  ``{DeepNIS: Deep neural network for nonlinear electromagnetic inverse
  scattering},'' \emph{IEEE Transactions on Antennas and Propagation}, vol.~67,
  no.~3, pp. 1819--1825, 2018.

\bibitem{guo2021complex}
L.~Guo, G.~Song, and H.~Wu, ``{Complex-Valued Pix2pix—Deep Neural Network for
  Nonlinear Electromagnetic Inverse Scattering},'' \emph{Electronics}, vol.~10,
  no.~6, p. 752, 2021.

\bibitem{wei2018deep}
Z.~Wei and X.~Chen, ``{Deep-Learning Schemes for Full-Wave Nonlinear Inverse
  Scattering Problems},'' \emph{IEEE Transactions on Geoscience and Remote
  Sensing}, 2018.

\bibitem{chen2019learning}
G.~Chen, P.~Shah, J.~Stang, and M.~Moghaddam, ``{Learning-assisted
  multimodality dielectric imaging},'' \emph{IEEE Transactions on Antennas and
  Propagation}, vol.~68, no.~3, pp. 2356--2369, 2019.

\bibitem{sanghvi2019embedding}
Y.~Sanghvi \emph{et~al.}, ``{Embedding deep learning in inverse scattering
  problems},'' \emph{IEEE Transactions on Computational Imaging}, vol.~6, pp.
  46--56, 2019.

\bibitem{lu2018beyond}
Y.~Lu, A.~Zhong, Q.~Li, and B.~Dong, ``{Beyond finite layer neural networks:
  Bridging deep architectures and numerical differential equations},'' in
  \emph{International Conference on Machine Learning}.\hskip 1em plus 0.5em
  minus 0.4em\relax PMLR, 2018, pp. 3276--3285.

\bibitem{haber2017stable}
E.~Haber and L.~Ruthotto, ``{Stable architectures for deep neural networks},''
  \emph{Inverse problems}, vol.~34, no.~1, p. 014004, 2017.

\bibitem{weinan2017proposal}
E.~Weinan, ``{A proposal on machine learning via dynamical systems},''
  \emph{Communications in Mathematics and Statistics}, vol.~5, no.~1, pp.
  1--11, 2017.

\bibitem{shan2021physics}
T.~Shan, X.~Song, R.~Guo, M.~Li, F.~Yang, and S.~Xu, ``{Physics-informed
  Supervised Residual Learning for Electromagnetic Modeling},'' in \emph{2021
  International Applied Computational Electromagnetics Society Symposium
  (ACES)}.\hskip 1em plus 0.5em minus 0.4em\relax IEEE, 2021, pp. 1--4.

\bibitem{xiao_dynamical_2018}
L.~Xiao, Y.~Bahri, J.~Sohl-Dickstein, S.~S. Schoenholz, and J.~Pennington,
  ``Dynamical {Isometry} and a {Mean} {Field} {Theory} of {CNNs}: {How} to
  {Train} 10,000-{Layer} {Vanilla} {Convolutional} {Neural} {Networks},'' Jul.
  2018.

\bibitem{tarnowski_dynamical_2019}
W.~Tarnowski, P.~Warchol, S.~Jastrzebski, J.~Tabor, and M.~Nowak, ``{Dynamical
  isometry is achieved in residual networks in a universal way for any
  activation function},'' in \emph{The 22nd {International} {Conference} on
  {Artificial} {Intelligence} and {Statistics}}.\hskip 1em plus 0.5em minus
  0.4em\relax PMLR, 2019.

\bibitem{long2018pde}
Z.~Long, Y.~Lu, X.~Ma, and B.~Dong, ``{Pde-net: Learning pdes from data},'' in
  \emph{International Conference on Machine Learning}.\hskip 1em plus 0.5em
  minus 0.4em\relax PMLR, 2018, pp. 3208--3216.

\bibitem{ruthotto2020deep}
L.~Ruthotto and E.~Haber, ``{Deep neural networks motivated by partial
  differential equations},'' \emph{Journal of Mathematical Imaging and Vision},
  vol.~62, no.~3, pp. 352--364, 2020.

\bibitem{he2016deep}
K.~He, X.~Zhang, S.~Ren, and J.~Sun, ``{Deep residual learning for image
  recognition},'' in \emph{Proceedings of the IEEE conference on computer
  vision and pattern recognition}, 2016, pp. 770--778.

\bibitem{clevert2015fast}
D.-A. Clevert, T.~Unterthiner, and S.~Hochreiter, ``{Fast and accurate deep
  network learning by exponential linear units (elus)},'' \emph{arXiv preprint
  arXiv:1511.07289}, 2015.

\bibitem{geffrin2005free}
J.-M. Geffrin, P.~Sabouroux, and C.~Eyraud, ``{Free space experimental
  scattering database continuation: experimental set-up and measurement
  precision},'' \emph{inverse Problems}, vol.~21, no.~6, p. S117, 2005.

\end{thebibliography}

% that's all folks
\end{document}